\RequirePackage{fix-cm}
\documentclass[smallcondensed]{svjour3}     % onecolumn (ditto)
\smartqed  % flush right qed marks, e.g. at end of proof
\usepackage{adjustbox}
\usepackage{textcomp}
\usepackage[numbered]{bookmark}
\usepackage{soul}
\usepackage{booktabs}
\usepackage{multirow}
\usepackage{float}
\usepackage{url}
\usepackage{tcolorbox}
\usepackage{graphicx}
\usepackage{textcomp}
\usepackage{xcolor}
\usepackage{url}
\usepackage{tabularx} 
\usepackage{balance}
\usepackage{stfloats}
\usepackage{fullpage}
\usepackage[justification=centering]{caption}
\usepackage{lineno}
\usepackage[title]{appendix}
\usepackage{ragged2e}

\DeclareCaptionType{Quote}
\makeatletter
\newcommand\subsubsubsection{\@startsection{paragraph}{4}{\z@}{-2.5ex\@plus -1ex \@minus -.25ex}{1.25ex \@plus .25ex}{\normalfont\normalsize}}
\newcommand\subsubsubsubsection{\@startsection{subparagraph}{5}{\z@}{-2.5ex\@plus -1ex \@minus -.25ex}{1.25ex \@plus .25ex}{\normalfont\normalsize}}
\makeatother

\newcommand{\eman}[1]{\textcolor{violet}{{\it [Eman says: #1]}}}

\newcommand{\RQA}{RQ$_1$: How have refactoring discussions on Stack Overflow grown over the years?}
\newcommand{\RQAA}{RQ$_{1.1}$: How have refactoring posts grown throughout the years?}
\newcommand{\RQAB}{RQ$_{1.2}$: What is the distribution of questions and answers among developers?}
\newcommand{\RQAC}{RQ$_{1.3}$: What are the tags that are associated with refactoring questions?}
\newcommand{\RQB}{RQ$_2$: What do developers discuss in refactoring based Stack Overflow posts?}
\newcommand{\RQBA}{RQ$_{2.1}$: What are the frequent terms utilized by developers in refactoring discussions?} 
\newcommand{\RQBB}{RQ$_{2.2}$: To what extent do traditional refactoring opportunities, known in existing literature, match with the challenges faced by developers in Stack Overflow posts?}
\newcommand{\RQBC}{RQ$_{2.3}$: What are the topics around software refactoring that are being asked by developers?} 
\newcommand{\RQC}{RQ$_3$: Which topics are the most popular and difficult among refactoring-related questions?} 

\begin{document}

\title{How Do I Refactor This? An Empirical Study on Refactoring Trends and Topics in Stack Overflow} 


\author{Anthony Peruma\thanks{Anthony Peruma is the corresponding author.} \and
        Steven Simmons\and
        Eman Abdullah AlOmar\and
        Christian D. Newman\and
        Mohamed Wiem Mkaouer\and
        Ali Ouni
}


\institute{Anthony Peruma, Steven Simmons, Eman Abdullah AlOmar, Christian D. Newman, Mohamed Wiem Mkaouer  \at
              Rochester Institute of Technology \\
              Rochester, New York, USA\\
              \email{axp6201@rit.edu, sds9278@rit.edu, eman.alomar@mail.rit.edu, cnewman@se.rit.edu, mwmvse@rit.edu}
           \and
           Ali Ouni \at
              ETS Montreal, University of Quebec \\
              Montreal, Quebec, Canada\\
              \email{ali.ouni@etsmtl.ca}
}

\date{Received: date / Accepted: date}

\maketitle

\begin{abstract}
An essential part of software maintenance and evolution, refactoring is performed by developers, regardless of technology or domain, to improve the internal quality of the system, and reduce its technical debt. However, choosing the appropriate refactoring strategy is not always straightforward, resulting in developers seeking assistance. Although research in refactoring is well-established, with several studies altering between the detection of refactoring opportunities and the recommendation of appropriate code changes, little is known about their adoption in practice. Analyzing the perception of developers is critical to understand better what developers consider to be problematic in their code and how they handle it. %This would feed back to ongoing refactoring research and orient future investigations towards the developer's interest areas.
Additionally, there is a need for bridging the gap between refactoring, as research, and its adoption in practice, by extracting common refactoring intents that are more suitable for what developers face in reality. In this study, we analyze refactoring discussions on Stack Overflow through a series of quantitative and qualitative experiments. Our results show that Stack Overflow is utilized by a diverse set of developers for refactoring assistance for a variety of technologies. Our observations show five areas that developers typically require help with refactoring-- Code Optimization, Tools and IDEs, Architecture and Design Patterns, Unit Testing, and Database. We envision our findings better bridge the support between traditional (or academic) aspects of refactoring and their real-world applicability, including better tool support.

% \and Topic Models \and Latent Dirichlet Allocation}
\end{abstract}

\section{Introduction}
\label{Section:Introduction}
Refactoring is a disciplined technique and a fundamental activity in software development. From the most structurally simplistic task of renaming an identifier to more complex structural changes such as extracting and moving a method, developers refactor their code to improve the internal quality of their systems while still preserving the system's behavior \cite{Fowler2018refactoring}. According to its definition, refactoring is applied to enforce best design practices or to cope with design defects. Thus, academic research has been recommending refactoring to either fix code and test smells (e.g., long methods, God classes, etc.) \cite{Fontana2015XP,Lambiase2020ICPC} or improve structural metrics (e.g., coupling, cohesion, cyclomatic complexity, etc.) \cite{DuBois2004WCRE,Chavez2017SBES}. However, recent surveys have shown the lack of adoption of refactoring tools in practice \cite{Murphy-Hill2012TSE,Kim2014TSE}, in part due to the lack of support for the types of problems developers face when refactoring. One potential problem is that while refactorings tend to be applied alongside other development and maintenance tasks \cite{pantiuchina2020developers,alomar2020howwe}, automated refactoring tools do not fully take this aspect into account. In addition, this interleaving of actions indicates that there may be some relationship between refactoring and other development tasks, meaning that these other tasks are essential for comprehending the refactoring. Unfortunately, there is still a lack of thorough understanding of the rationale behind applying a refactoring or the problems developers face when determining which refactorings to apply and how to apply them in practice.

Prior studies that have examined the rationale behind the application of refactoring activities relied on the examination of project artifacts such as commit logs (focusing mainly on a selected set of open-source projects) \cite{Peruma2018IWOR,Peruma2019MobileSoft,AlOmar2019IWoR,Peruma2020JSS} or interviewing a selected set of developers \cite{Kim2014TSE,Danilo2016FSE}. However, interviews are a limited resource and may not fully generalize due to the necessarily limited number of people they gather information from. In addition, the natural language analysis of commit logs is not always a feasible approach because developers typically use high-level descriptions; rarely mentioning specific refactorings or project-specific reasoning for why changes have been made beyond high-level reasoning, such as improving readability/comprehension of the code, eliminating code/design smells, and fixing defects \cite{Peruma2018IWOR,Peruma2020JSS,Murphy-Hill2012TSE}. As a result, prior work has highlighted the challenges in correlating rationale obtained from various software artifacts with refactoring activities \cite{Peruma2018IWOR,Peruma2020JSS,Murphy-Hill2012TSE}. For instance, while we know that developers refactor to improve readability, it is still unclear how a developer identifies a readability issue, how the developer plans to perform this improvement, and how we can accurately obtain this information from software artifacts that are updated in response to these changes.

There are other sources of information, besides commit messages and interviews, for understanding the rationale behind refactoring; the most popular programming-specific question and answer forum on the internet, Stack Overflow, is one of them. With over 19 million questions \cite{stackoverflowQuestions} and one million users  (as of January 2019) \cite{stackoverflowUsers}, Stack Overflow provides developers with a mechanism to seek advice and help from the community on a wide range of software development topics. Hence, through this study, we are presented with a diverse and representative view of developer discussions around refactoring; thereby, gaining a real-world understanding of not only the volume of discussions around refactoring but also the topics that are of most interest and challenging to developers. 

\subsection{Motivation}

%Hence, to better understand the current challenges of software refactoring in practice, it is crucial to investigate developers discussions in Q\&A forums.
The field of software refactoring research is continually evolving. For instance, the initial definition of refactoring indicates that the driver to refactor source code arises from the degradation of the internal quality of a system \cite{MensTSE2004}. However, through empirical \cite{alomar2020toward} and developer-based \cite{Danilo2016FSE} studies, it is apparent that code smells may not be the only reason developers apply refactorings. Given that it is common for academic research to recommend refactoring based on the presence of code smells, there is a clear need for a stronger understanding of the other drivers of refactorings so that appropriate recommendations can be made in those contexts as well. Quote \ref{Quote:Motivate04_} is an example of a developer seeking help with a code change task that was caused as a consequence of several factors: high complexity metrics and poor readability. They found some advice pointing at the Strategy design pattern as a solution, but are unsure of whether it is overkill for this problem or not. In this context, several studies have proposed solutions to fix brain methods or reduce cyclomatic complexity. However, little is known on how to address a problem that contains both of these characteristics. Therefore, there is a need to investigate techniques that can approach these types of issues holistically.
%by Real-world systems are complex and have usually undergone multiple rounds of evolution, making refactoring a challenging task. Because our understanding of how developers comprehend and apply the notion of refactoring is limited, it is difficult to understand how best to support them in their refactoring tasks. 
Prior work has tried to fill this gap by interviewing and surveying developers. However, these studies are still limited in how much they can tell us about the general population of developers. The question is: \textit{Are developers applying refactorings in the same environments, on problems with the same characteristics and context, as researchers assume?} To answer this question, we need to understand refactoring as it is discussed in day-to-day development activities. Thus, we focus our attention on real-world challenges developers currently face when refactoring their systems. For instance, consider the example in Quote \ref{Quote:Motivate02_}. The majority of refactoring-related research focuses on statically typed programming languages, such as Java. However, dynamically typed languages such as Python and other software artifacts such as databases and configuration files may also need to be part of some refactorings, such as the rename refactoring.

%are gaining popularity in the industry, and they too require refactoring support. Furthermore, this example also highlights the coupling between multiple project artifacts, which current research does not account for, and the need for automated synchronization of refactoring operations between related artifacts.

\begin{center}
\fbox{\parbox{\dimexpr\linewidth-2\fboxsep-2\fboxrule\relax}{\centering
\textbf{Refactoring switch statement for Data to different types of data}
\begin{flushleft}
\textit{``My mission is to refactor a switch statement that was poorly written (it makes the cyclomatic complexity spike). In short, there is a class that parses a file for various values. I have read up on the topic and while there seems to be plenty of help, it all seems to be pointing at the Strategy design pattern, which (I believe) would overkill my problem. In my project, there are multiple classes like this, with some of them having upwards of 25 case statements."}
\end{flushleft}
}}
\captionof{Quote}{A developer runs into challenges trying to improve the quality metrics of the code.
\cite{motivate04}.\label{Quote:Motivate04_}}
\end{center}

\begin{center}
\fbox{\parbox{\dimexpr\linewidth-2\fboxsep-2\fboxrule\relax}{\centering
\textbf{What refactoring tools do you use for Python?}
\begin{flushleft}
\textit{``I have a bunch of classes I want to rename. Some of them have names that are small and that name is reused in other class names, where I don't want that name changed. Most of this lives in Python code, but we also have some XML code that references class names... Does anyone have experience with Python refactoring tools ? Bonus points if they can fix class names in the XML documents too."}
\end{flushleft}
}}
\captionof{Quote}{An example of a refactoring challenge a developer faces when refactoring Python code.
\cite{motivate02}.\label{Quote:Motivate02_}}
\end{center}

% \begin{center}
% \fbox{\parbox{\dimexpr\linewidth-2\fboxsep-2\fboxrule\relax}{\centering
% \textbf{How to simplify a switch statement to reduce the cyclomatic complexity?}
% \begin{flushleft}
% I have a big switch statement that has a cyclomatic complexity of 31 and it must be refactorized to at least 25...Here is the code...
% \end{flushleft}
% }}
% \captionof{Quote}{An example of a refactoring challenge a developer faces when refactoring Python code.
% \cite{motivate03}.\label{Quote:Motivate03_}}
% \end{center}

As one of the premier online question and answer resources for developers, Stack Overflow is an ideal candidate to mine for real-world discussions around programming-related challenges. The diversity of the user base presents us with an opportunity to study what and how developers refactor their systems effectively. The content we mine from Stack Overflow will assist us in gauging the extent to which current research in refactoring is lacking and understand the reasons for gaps between the application of automated refactoring and the way developers perceive it \cite{Danilo2016FSE}. Furthermore, as this study captures the state of real-world refactoring at this current point in time, it becomes an ideal candidate for future replication-based studies to determine the extent to which real-world refactoring has evolved, such as the technologies and challenges developers face when refactoring their systems.

\subsection{Goal \& Research Questions}
%The goal of this study is to \textit{understand the trends and challenges around developer discussions on software refactoring concepts and activities}. Our study helps facilitate improvements in software refactoring activities. We envision our findings used as input by educators \ali{may be also practitioners, researchers, junior and senior developers, newcomers to a given project etc.} to better prepare developers on the common and essential areas of refactoring that they would face in real-world settings. Additionally, IDE and tool vendors will find our results useful in improving their instruments to better support developers. 
The goal of this study is to \textit{understand the trends and challenges around developer discussions on software refactoring concepts and activities}. We envision our findings used as input by practitioners, researchers, and educators in understanding the current state of refactoring trends and challenges and determine the extent to which traditional (or academic) viewpoints of refactoring need revising based on real-world applicability. Additionally, IDE and tool vendors that offer refactoring support will find our results helpful in improving their instruments to better support developers.
Hence, we first start by examining the volume of refactoring posts and user contributions. Through this examination, we reveal how popular and valuable Stack Overflow is in the developer community in discussing refactoring topics. Once we establish this volume, our next task is to determine the topics of discussion. We achieve this in three steps by analyzing the natural language text in the body of a question. We first identify common phrases; then, we look for specific refactoring terminology in the body. Finally, our third step is grouping similar questions based on the terms they utilize and then determining the category (or topic) of these groupings. The final analysis of this study utilizes the identified categories to determine the type of questions that are challenging to answer. Thus, we define and address the following research questions:
%Hence, in this study, we first start by examining the volume of refactoring posts and user contributions. Through this examination, we gain a sense of understanding of how popular and useful is Stack Overflow in the developer community in discussing refactoring topics. Once we establish this volume, our next task is to determine the topics of discussions. We achieve this in three steps. Our first step examines the tags. By studying this fixed set of terms, we understand the specific areas and technologies that represent the posts. However, as these results are too high-level, we go deeper by analyzing the textual content in posts identifying common phrases. However, these common phrases and terms can be vast due to the volume of data in our study. Hence, our third step is the grouping of similar questions based on the terms they utilize and then determining the category (or topic) these groupings. The final analysis of this study utilizes the identified categories to determine the type of questions that are challenging to answer. Thus, we define and address the following research questions:

%By decomposing our overall goal, we intend to : (1) understand the growth rate and trends of refactoring discussions, (2) discover the topics around refactoring discussions, and (3) quantitatively investigate attributes associated with refactoring discussions. 

\textbf{\RQA}
Through this research question, we gain insights into the growth in refactoring related discussions throughout the years. This question explores the volume of questions and answers on Stack Overflow, along with the community members responsible for creating these posts. Additionally, we examine the tags that accompany refactoring questions, which presents us with high-level information into the areas that involve refactoring.
%Through this research question, we gain insight into the growth in refactoring related discussions throughout the years. This research question will help us determine the magnitude to which developers face challenges, seek advice, and share knowledge around refactoring. 
%\textbf{\RQB}
%This research question helps us understand if a select set of Stack Overflow users are responsible for a majority of refactoring related questions and answers. Furthermore, we aim to learn the level of segregation among the Stack Overflow users base with regards to the level of involvement in refactoring based discussions.
%\textbf{\RQC}
%Tags are an integral part of Stack Overflow questions as they represent the subject of the question in a single word. Using tags, developers can locate questions of interest that they can contribute to. From this research question, we gain insight into the tags that accompany refactoring questions, which presents us with high-level information into the areas that involve refactoring.

\textbf{\RQB}
This research question utilizes natural language techniques to identify the key terms and phrases developers utilize in crafting refactoring questions. Our analysis presents a real-world and granular view of the problematic or challenging software refactoring areas developers require assistance.
%\textbf{\RQD}\eman{since the answer to this question involves refactoring operation-related discussion, I suggest to either rephrase it or explicitly include the term "refactoring operation" in the question} \ali{Agree}
%As a question and answer site, the fundamental purpose of Stack Overflow is to facilitate developers to utilize natural language to either describe the problem they require help with or provide advice/solutions to these questions. This research question identifies the key terms utilized by developers in these posts to point us in the areas associated with real-world refactoring discussions.

%\textbf{\RQE}
%Building on the previous research question, this question utilizes natural language techniques to perform a deep dive into the questions asked by developers. From this research question, we present a real-world and granular view of the problematic or challenging areas of software refactoring that developers require assistance with.

\textbf{\RQC}
This research question presents us with the opportunity to understand the refactoring topics that are popular among developers and topics that are difficult to answer on Stack Overflow. Additionally, we also examine unanswered questions. %By knowing the popular and challenging topics, researchers and educators will be better equipped to prioritize areas of refactoring that are of most interest and challenging to developers and spurn research into improving developer productivity.

\subsection{Contributions}
This study provides the community of researchers and practitioners with insights on the discussion of refactoring on Stack Overflow. More specifically, are contributions are outlined below:
\begin{itemize}
    \item Mining and extraction of 9,489 Stack Overflow questions related to refactoring.
    \item A series of quantitative and qualitative experiments on the extracted questions to show the growth and trends of refactoring discussions between developers, such as:
    \begin{itemize}
    \item The frequently occurring set of tags and terminology in questions, which also include an analysis of the use of refactoring specific terminology.
    \item The primary topics for refactoring questions, including the popular and challenging topics, along with an analysis of unanswered questions.
    \end{itemize}
\end{itemize}
We also make available our dataset for replication and extension purposes \cite{ProjectWebsite}.

\section{Related Work}
\label{Section:RelatedWork}

Before discussing the research works related to our study, we first present a brief overview of the state of research in the field. Since developers discuss multiple aspects of refactoring on StackOverflow, we believe it is essential that readers comprehend the field's evolution to understand the extent to which academia address practical challenges developer face. The spectrum of research exploring the practice of refactoring covers a wide variety of dimensions. One of the earliest studies, by Mens and Tourwe \cite{MensTSE2004}, provides an overview of existing research in the field of software refactoring. They discuss the existing literature in terms of refactoring activities and techniques, refactoring tool support, and the impact of refactoring on the software process. Further studies on refactoring focus on studying the impact of refactoring on quality (e.g., \cite{moser2006does,bavota2015experimental,alomar2019impact,pantiuchina2018improving,cedrim2016does,wilking2007empirical}), identifying refactoring opportunities (e.g., \cite{fontana2012automatic,palomba2013detecting}), recommending refactoring operations (e.g., \cite{mkaouer2015many,bavota2014recommending,ouni2016multi}), and implementing refactoring tools (e.g., \cite{roberts1997refactoring,mazinanian2016jdeodorant,kim2010ref,tsantalis2018accurate,silva2020refdiff,moghadam2021refdetect}). 

As our study is around developer discussions on refactoring, our discussion of related work is limited to studies investigating the rationale and motivations of why developers refactor code or research into discussions around tools/technologies that developers rely on for refactoring code. To this extent, we group our set of related works into two groups. The first group contains studies that mine Stack Overflow posts and report on refactoring-based discussions among developers. The second group of studies is based on developer interviews or analysis of project artifacts to understand developer refactorings.

%While there are numerous studies based on (1) mining Stack Overflow posts and (2) refactoring, our analysis of related works is limited to studies investigating the rationale and motivations as to why developers refactor code or research into discussions around tools/technologies that developers rely on for refactoring code. To this extent, we group our set of related works into two groups. The first group contains studies that mine Stack Overflow posts and report on specific aspects of refactorings based discussions among developers. The second group of studies is either based on developer interviews or analysis of project artifacts to understand developer refactorings. 

\subsection{Refactoring Related Discussions in Stack Overflow Posts}
Pinto and Kamei \cite{Pinto20137WRT} mine Stack Overflow posts to study discussions around refactoring tools. From a tool features perspective, the authors observe that developers prefer tools that provide refactoring recommendations and support for database and multi-language refactoring. Additionally, findings from this study show that in addition to usability issues, the lack of trust in tools is one of the major barriers to adoption. In their study on code smells and antipattern discussions on Stack Overflow, Tahir et al. \cite{Tahir2018EASE} observe that the majority of answers to these questions did not provide refactoring recommendations; instead, the answers provide details around the code smell/antipattern. Furthermore, developers do not frequently refer to refactoring operations by name in the posts and refer to some design patterns as potential refactoring solutions. In a subsequent study \cite{tahir2020large}, the authors performed a large-scale study that explores how developers discuss code smells and antipatterns in Stack Exchange. The authors show that most of the questions focus on the following code smells: Duplicated Code, Spaghetti Code, God Class, and Data Class. As for the programming languages, most of the discussions focus on popular languages like C\#, JavaScript, and Java. Although Java has greater tooling support, other platforms such as C\# and JavaScript are lacking in support. Findings by Tian et al. \cite{Tian2019ICSA}, from their study on architecture smells, show the lack of tools for refactoring architecture smells. The authors also highlight that even though there exist specialized tools to refactor architecture smells, these tools are not mentioned in the Stack Overflow posts; instead, developers mention the use of common code smell detection tools. Additionally, the authors also observe that time and costs involved with detecting and refactoring architecture are very concerning for most developers. In a preliminary study specializing in the refactoring of Java 8 streams for parallelization, Tang et al. mined Stack Overflow posts for discussions around Java 8 streams \cite{Tang2018ICSE}. As part of their findings, the authors mention that 5\% of questions around Java 8 streams remain unanswered. A preliminary study by Choi et al. \cite{Choi2015IWSC} on code clones shows that most discussions are related to refactoring with the need for more support for clone refactoring tools. Openja et al. \cite{Openja2020ICSME} utilize topic modeling to study release engineering questions. More specifically, the authors examine popular topics among developers and those that are difficult to answer. In this study, the authors identify 38 topics from which questions around security are both challenging and popular.% We summarize these state-of-the-art studies in Table \ref{Table:RelatedWork_SO}.

While our study also mines Stack Overflow posts, the key difference between our study and the works described above is that our study is not limited to a specific programming language, paradigm, or technology. We retrieve and analyze any post related to the concept of refactoring.

% \begin{table*}
%   \centering
% 	 \caption{A summary of the literature of refactoring related discussions in Stack Overflow posts.}
% 	 \label{Table:RelatedWork_SO}
% \begin{adjustbox}{width=1.0\textwidth,center}
% %\begin{adjustbox}{width=\textheight,totalheight=\textwidth,keepaspectratio}
% \begin{tabular}{llllllll}\hline
% \toprule
% \bfseries Study & \bfseries Year & \bfseries Focus &  \bfseries No of post &\bfseries Main finding    \\
% \midrule
% Pinto and Kamei \cite{Pinto20137WRT} & 2013 & Refactoring tools & 1,439 & Practitioners do not often rely on the
% existing refactoring tools\\
% Choi et al. \cite{Choi2015IWSC} & 2015 & Code clone &  925 & Need
% for basic support of development of tools for clone refactoring \\
% Tahir et al. \cite{Tahir2018EASE} & 2018 & Code smells \& anti-patterns & 3,109 & Need for more context-based evaluations of code smells and anti-patterns \\
% Tian et al. \cite{Tian2019ICSA} & 2019 & Architecture smells & 207 & Lack of tools for detecting and refactoring architecture smells\\
% Tahir et al. \cite{tahir2020large} & 2020 & Code smells \& anti-patterns & 4,036 & Need for a unified, constantly updated, catalog of smells and anti-patterns \\
% Openja et al. \cite{Openja2020ICSME} & 2020 & Release engineering & 260,023  & 
%  `Security' is both popular and difficult release engineering topic  \\
% \bottomrule
% \end{tabular}
% \end{adjustbox}
% %\end{sideways}
% \end{table*}

\subsection{Refactoring} 
Arnaoudova et al.  \cite{Arnaoudova2014TSE} surveyed 71 developers to understand the importance of rename refactoring operations. The findings show that developers consider renaming refactoring a challenging activity, and they frequently perform rename operations on the source code. In their study of utilizing the commit log to contextualize renaming operations, Peruma et al. \cite{Peruma2018IWOR} observe developers perform renames as part of addressing defects and unit tests. The authors also highlight that the commit log alone cannot be utilized to gain insights into renaming operations. In a preliminary study on the refactorings of Android apps, Peruma \cite{Peruma2019MobileSoft} performs a topic modeling analysis on the commit log. The author shows that developers refactor apps for reasons such as improving code readability, fixing defects, and enhancing the app's design. A survey with developers at Microsoft by Kim et al. \cite{Kim2014TSE} shows that there are costs and risks involved with the performance of refactoring activities and also the need for more tool-based support. Furthermore, the survey results show that developers do not consider refactoring to be confined to only behavior preserving transformations; this is in contrast to the academic definition. In their study, Danilo et al. \cite{Danilo2016FSE} identify that changes in requirements are one of the key reasons that drive developers to refactor code. The authors also show that `Extract Method' is a frequently occurring refactoring operation and identify 11 motivations for applying this operation. 
Additionally, the authors also identify that developers are concerned about introducing duplicate code. Finally, the authors also indicate that developers more frequently apply refactorings manually than using a tool. They also report that a lack of trust in tools is a key concern among developers. Murphy-Hill et al. \cite{Murphy-Hill2012TSE} note that the majority of refactoring operations are performed manually. However, the rename operation is frequently performed using a tool. Additionally, the authors also indicate that there exist instances where developers utilize tools to perform refactorings in batches. Examining commit messages, the authors show that it is not feasible to determine if a refactoring was applied based on the message in the commit log.

More recently, Pantiuchina et al. \cite{pantiuchina2020developers} present a mining-based study to investigate why developers perform refactorings by analyzing the history of 150 open-source systems. Particularly, they analyze 551 pull requests containing refactoring operations and produce a refactoring taxonomy that generalizes existing literature. In a large-scale empirical study on refactoring, AlOmar et al. \cite{alomar2020howwe} explore what motivates developers to apply refactorings by mining and automatically classifying a set of 111,884 commits containing refactoring activities extracted from 800 open-source Java projects. Their findings show that fixing code smells is not the main driver for developers to refactor their code. Developers refactoring for various reasons (e.g., feature addition, bugfix), going beyond its traditional definition. Furthermore, recent studies \cite{alomar2019impact,pantiuchina2018improving} show that there is a misperception between the state-of-the-art structural metrics widely used as indicators for refactoring and what developers consider to be an improvement in their source code. The authors identified (among software quality models) metrics that align with the vision of developers on the quality attributes they explicitly state they want to improve. A number of studies have recently focused on the documentation of refactoring. AlOmar et al. \cite{AlOmar2019IWoR,alomar2020toward,alomar2020howwe} have explored how developers document their refactoring activities in commit messages; this activity is called Self-Affirmed Refactoring (SAR). In particular,  SAR indicates developers' explicit documentation of refactoring operations intentionally introduced during a code change. Based on their empirical investigation, developers tend to use a variety of textual patterns to document their refactoring activities, besides '\textit{refactor}', such as '\textit{redesign}', '\textit{reorganize}', and '\textit{polish}'. These patterns can be either (1) generic, providing a high-level description of the refactoring, or (2) specific by explicitly mentioning the rationale behind the applied refactoring operations.

\section{Study Methodology \& Dataset Construction}
\label{Section:Dataset}

%\eman{I found the methodology section to be brief and it does not reflect the efforts made. Beside adding an approach for each RQ, I am suggesting to give a big picture here in this section. I am suggesting to frame this section as follows if this is the approach we are doing: (1) Download SO dump, (2) Identify refactoring tags, (3) Extract refactoring posts, (4) Identify refactoring topics } \anthony{updated the paragraph} \ali{maybe Figure 1 could be also updated to reflect these steps 1 to 4 explicitly so that the reader can match the current text with the figure} \mohamed{Agree}

Our research methodology follows a mixed-methods approach, where we collect and analyze both quantitative and qualitative data \cite{Tashakkori1998Mixed}. This approach provides insight into relationships between qualitative and quantitative data and allows us to present representative samples from the dataset to complement our findings. More specifically, our approach utilizes well-established statistical measures, including unsupervised machine learning and natural language processing techniques which we apply to our dataset to report on trends and patterns of refactoring posts. Additionally, we also manually review a statistically significant sample of refactoring posts (body and metadata) to gain further insight into developers' refactoring challenges to supplement and correlate our quantitative findings.

\begin{figure}[h]
 	\centering
 	\includegraphics[trim=0cm 0cm 0cm 0cm, width=1\linewidth]{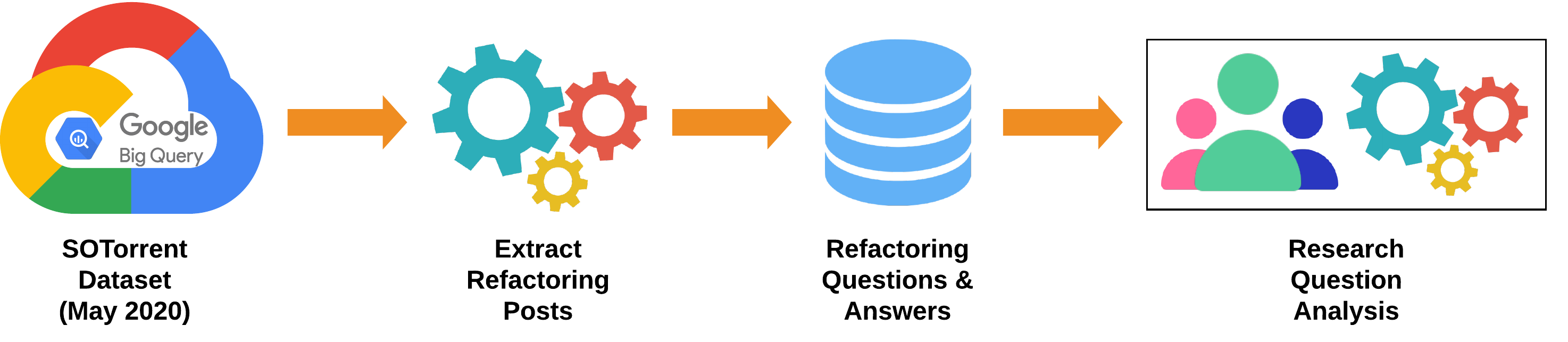}
 	\caption{Overview of our study methodology.}
 	\label{Figure:diagram_methodology}
\end{figure}

Figure \ref{Figure:diagram_methodology} outlines the methodology for our study.  Our methodology involves three key activities-- (1) obtaining a recent and representative Stack Overflow dataset, (2) identifying and extracting refactoring posts, (3) analyzing the refactoring posts (textual content and metadata). In summary, our study utilizes \textit{SOTorrent} \cite{Baltes2018MSR} to obtain Stack Overflow posts. From this dataset, we extract all refactoring posts (based on the tag and title of the post) and analyze these discussion posts via both manual inspection and automated mechanisms to help answer our research questions. In the following subsections, we describe in detail the elements and activities that were part of our methodology. Furthermore, we have the complete dataset available on our website for replication and extension purposes \cite{ProjectWebsite}.

%[ORIGINAL] The key to our study is obtaining a recent and representative dataset. As such, the methodology for our study involves three key activities-- (1) obtaining a recent and representative Stack Overflow dataset, (2) identifying and extracting refactoring posts, (3) analyzing the refactoring posts (textual content and metadata). As depicted in Figure \ref{Figure:diagram_methodology},  our study utilizes \textit{SOTorrent} \cite{Baltes2018MSR} to extract refactoring related discussions. Next, analyze these discussion posts via both manual inspection and automated mechanisms to help answer our research questions. In the following subsections, we describe in detail the elements and activities that were part of our data collection methodology. Furthermore, we have the complete dataset available on our website for replication and extension purposes \cite{ProjectWebsite}. 

\subsection{SOTorrent}
\label{Section:Methodology_SOTorrent}
For our study, we utilized the March 2020 release of the \textit{SOTorrent} dataset available on Google BigQuery\footnote{\url{https://bigquery.cloud.google.com/dataset/sotorrent-org:2020_03_15}}. The \textit{SOTorrent} dataset is constructed using a data dump provided by Stack Overflow. The dataset contains the version history of each post along with other relevant metadata such as the author of the post, score, view count, answer count, etc. Provided below, we briefly describe the attributes in the dataset that we utilized in our experiments.
\begin{itemize}
    \item \textbf{Posts}: There are three types of posts -- (1) question, (2) answer, and (3) accepted answer. A question post contains details about the problem/challenge faced by the developer. A question consists of a title and body field; the title is a plain text field, while the body supports a limited set of formatting. Responses to a question can fall into one of two categories: an accepted answer or a non-accepted answer\footnote{\url{https://stackoverflow.com/help/accepted-answer}}. A response post contains only a body field. While a question can have only one accepted answer, it can have more than one non-accepted answer. An accepted answer indicates that the developer asking the question considers this specific answer as a solution to the question asked. In other words, an accepted answer acts as a means of informing the community that the specific answer was the intended solution for the question. Only the developer asking the question can mark an answer as an acceptable answer.
    
    \item \textbf{Tags}: As part of creating a question, the developer needs to associate the question with one to five tags-- a word or phrase that describes the question\footnote{\url{https://stackoverflow.com/help/tagging}}. Only questions can be associated with tags. Tags permit site users to access a particular set of questions that is of interest to them. Stack Overflow discourages the creation of arbitrary tags and instead recommends the use of predefined tags.
    
    \item \textbf{Score}: Associated with posts, this metric is based on the Upvotes the post receives. The higher the score value, the more useful the post to the community\footnote{\url{https://stackoverflow.com/help/privileges/vote-up}}.
    
    \item \textbf{View Count}: Associated with only questions, this metric corresponds to the number of times the post was viewed\footnote{\url{https://meta.stackexchange.com/questions/90187}}.
    
    \item \textbf{Favorite Count}: Associated with only questions, this metric indicates the number of times site users have marked the post as a favorite.
\end{itemize}

\subsection{Refactoring Posts}
\label{Section:Methodology_Posts}
Even though \textit{SOTorrent} contained all Stack Overflow posts, our research focused on refactoring related discussions. To this end, we performed an extraction of relevant question-based posts from \textit{SOTorrent}. Our process involved retrieving questions that had either been tagged with a word containing the term \textit{`refactor'} or contained the term in the title. For each extracted question, we also extracted all answer-based posts  (including accepted answers) associated with the question. We excluded searching for the term \textit{`refactor'} in the body of the post as we observe that such an action leads to an increase of false positives in our dataset. Looking at such posts, we observe developers mentioning the term \textit{`refactor'} in passing, even though the post is not about the actual refactoring of code. For instance, in question-- Quote \ref{Quote:Dataset_01}, the developer requests help to solve a runtime exception without refactoring the code. Additionally, in some instances, the developer refactors the code to make it easier to comprehend the problem \cite{dataset_02}. Furthermore, as mentioned by Rosen and Shihab \cite{Rosen2016ESE}, the title succinctly describes the primary purpose of the question.

%\ali{how about other terms: restructure, reorganize, move, improve, rename, etc. Eman already found different terms being used by developers. This issue could be at least discussed in the threats to validity section (precision vs recall of the keyword 'refactor')}\eman{I agree, it is important to highlight it in the threat}. \steven{Added this to the threats. Included an explanation for the rationale behind the choice that I recalled}

\begin{center}
\fbox{\parbox{\dimexpr\linewidth-2\fboxsep-2\fboxrule\relax}{\centering
\textit{``The problem I have is that for a series of incoming messages...which leads to exceptions within the DLL...Is there any way around this, given that refactoring the DLL isn't an option.''}
}}
\captionof{Quote}{A false positive refactoring example \cite{dataset_01}\label{Quote:Dataset_01}.}
\end{center}

\subsection{Results Analysis}
\label{Section:Methodology_Analysis}
\begin{figure}[h]
 	\centering
 	\includegraphics[trim=0cm 0cm 0cm 0cm, width=1\linewidth]{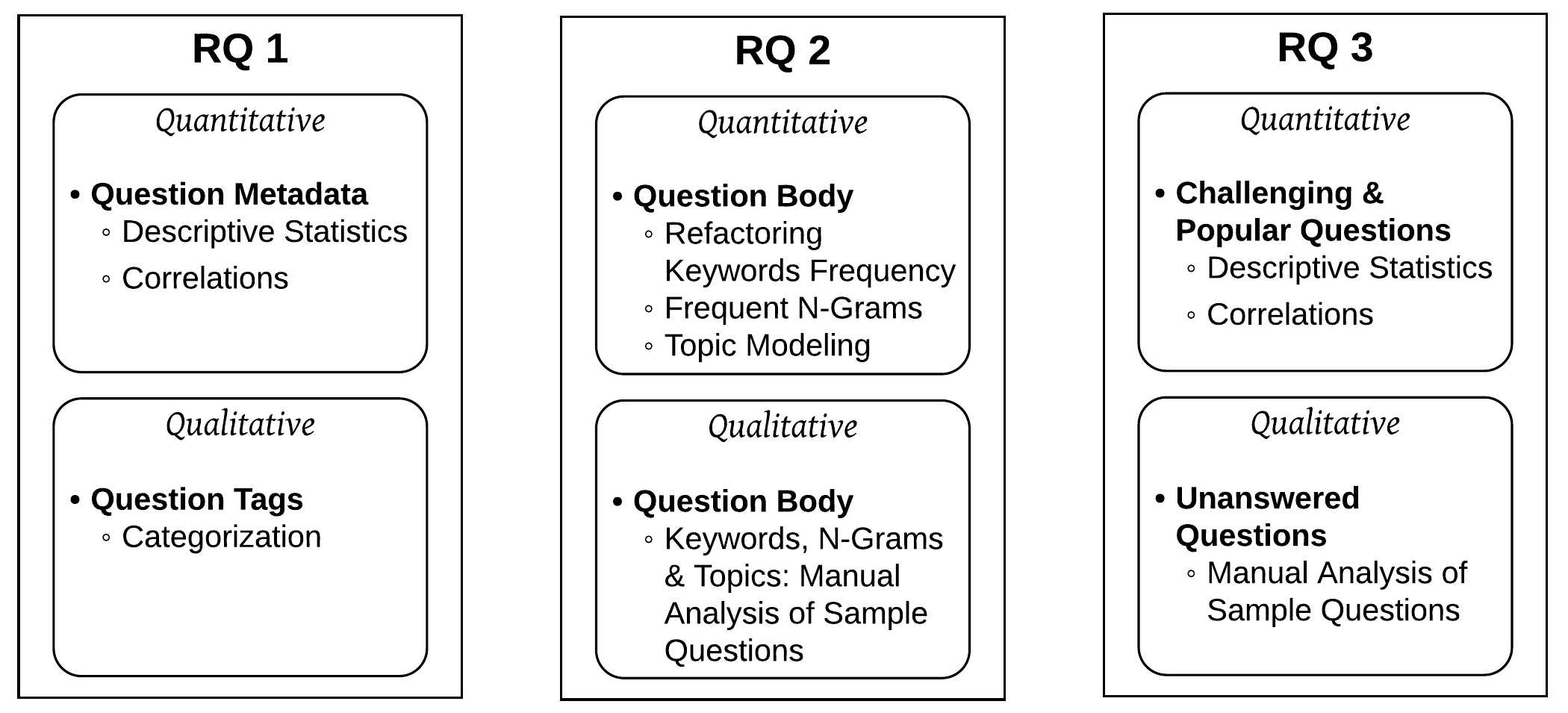}
 	\caption{Summary of the research approach we utilize to answer each research question (RQ).}
 	\label{Figure:diagram_methodologyRQs}
\end{figure}

Our study of refactoring posts includes a quantitative and qualitative approach. Our quantitative approach involves executing database queries and custom code/scripts, including a topic modeling analysis using an unsupervised machine learning algorithm. In the qualitative approach, two or more of the authors manually analyzed a statistically significant sample set of the data. Depicted in Figure \ref{Figure:diagram_methodologyRQs}, we summarize the type of research approach we utilize to answer each research question and the data on which it is applied. In Section \ref{Section:Results}, we elaborate in detail on our analysis approach to answering each research question.

% \subsection{Replication Package}
% We provide our comprehensive dataset on our website for replication and extension purposes \cite{ProjectWebsite}.

\section{Empirical Study Design \& Results}
\label{Section:Results}
In this section, we report on the results of our experiments. Our analysis contains three Research Questions (RQs), with certain RQs comprising of sub-RQs. In RQ$_1$, we look at the growth of refactoring posts, the individuals that ask and respond to questions, respectively, and the tags associated with questions. RQ$_2$ looks at identifying the rationale behind the refactoring posts. Specifically, the RQ examines the frequently occurring phrases in the body of a post and groups questions into categories (or topics) based on the textual content of the question. Finally, in RQ$_3$, we utilize the results from the prior RQ to identify popular and challenging topics.

For each RQ, we first explain the primary motivation(s) and the approach we undertake to produce the results; then, we present our findings. Further, where applicable, we also include examples to posts on Stack Overflow to provide the reader with more clarity on our observations. Finally, even though some tables and figures in the RQs show a subset of the data, we have the complete dataset available on our website for replication and extension purposes \cite{ProjectWebsite}. 

\subsection{\textbf{\RQA}}
\label{SubSection:RQA}
This RQ is composed of three sub-RQs exploring the growth of refactoring posts on Stack Overflow over the years. As a question and answer site, RQ$_{1.1}$ examines the growth of refactoring-based questions and answers on the site. In RQ$_{1.2}$, we examine if a selected set of Stack Overflow users is responsible for the majority of refactoring-related questions and answers. Finally, in RQ$_{1.3}$, we look at the tags used by developers in creating refactoring questions and the growth of these tags throughout the years.

\subsubsection{\RQAA}
\subsubsection*{\textit{Motivation \& Approach:}}
In this sub-RQ, we examine the growth of refactoring questions on Stack Overflow. The purpose of this sub-RQ is to understand the extent to which developers require help and advice on refactoring related problems and how often they receive the assistance they seek. To do this, we extract all questions that had the term `refactor' in either the title or tag. Next, for each question, we extracted all answers (i.e., accepted and non-accepted) associated with the question.

\subsubsection*{\textit{Findings:}}
In total, we extracted 9,489 refactoring-related questions, from which, 828 ($\sim$8.73\%) of the questions did not have an associated answer. Regarding accepted answers, 6,112 ($\sim$64.41\%) of the questions had an accepted answer. Table \ref{Table:DataCollection} provides a summary of the collected data. 

\begin{table}
\centering
\caption{Summary of collected data.}
\label{Table:DataCollection}
\begin{tabular}{@{}lr@{}}
\toprule
\multicolumn{1}{c}{\textbf{Item}} & \multicolumn{1}{c}{\textbf{Count}} \\ \midrule
Number of questions & 9,489 \\
Number of accepted answers & 6,112 \\
Number of non-accepted answers & 14,491 \\
Number of questions without an answer & 828 \\
Number of questions with a `refactor' tag & 7,024 \\
Average number of answers per question & 2 \\
Average number of tags per question & 3 \\ \bottomrule
\end{tabular}
\end{table}

\begin{figure}
 	\centering
 	\includegraphics[trim=2.3cm 1.5cm 3.0cm 0cm,width=1\linewidth]{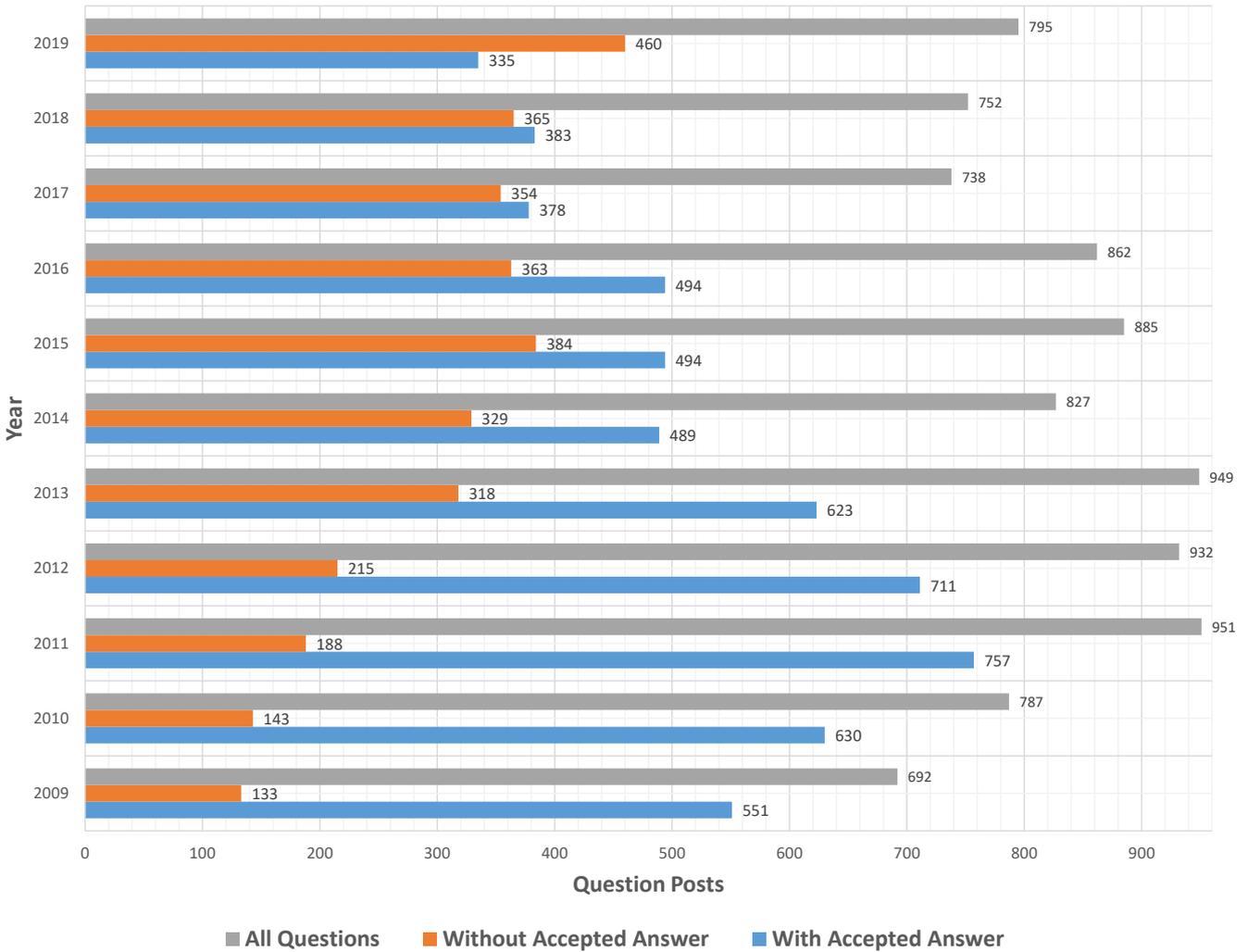}
 	\caption{Count of refactoring questions and accepted answers, per year.} 
 	\label{Figure:chart_question_withwithoutanswer}
\end{figure}

Figure \ref{Figure:chart_question_withwithoutanswer} shows a yearly breakdown of refactoring questions with and without an accepted answer. In this chart, for each year, the orange bars are questions without an accepted answer, while the blue bars are questions with an accepted answer. The gray bars represent all questions (i.e., questions with and without an accepted answer). Our count of questions with an accepted answer is constrained to question-accepted answer pairs created in the same year. It should be noted that Stack Overflow was launched in September 2008, and our dataset contains posts up to March 2020; hence our analysis excludes posts created in these two years. A first glance at this chart shows that, other than for the year 2019, the number of questions with an accepted answer outnumber questions without an accepted answer. However, also shown in this chart is that as the years progress, the number of questions with accepted answers decreases while questions without accepted answers increases.

\begin{figure}
 	\centering
 	\includegraphics[width=1.0\linewidth]{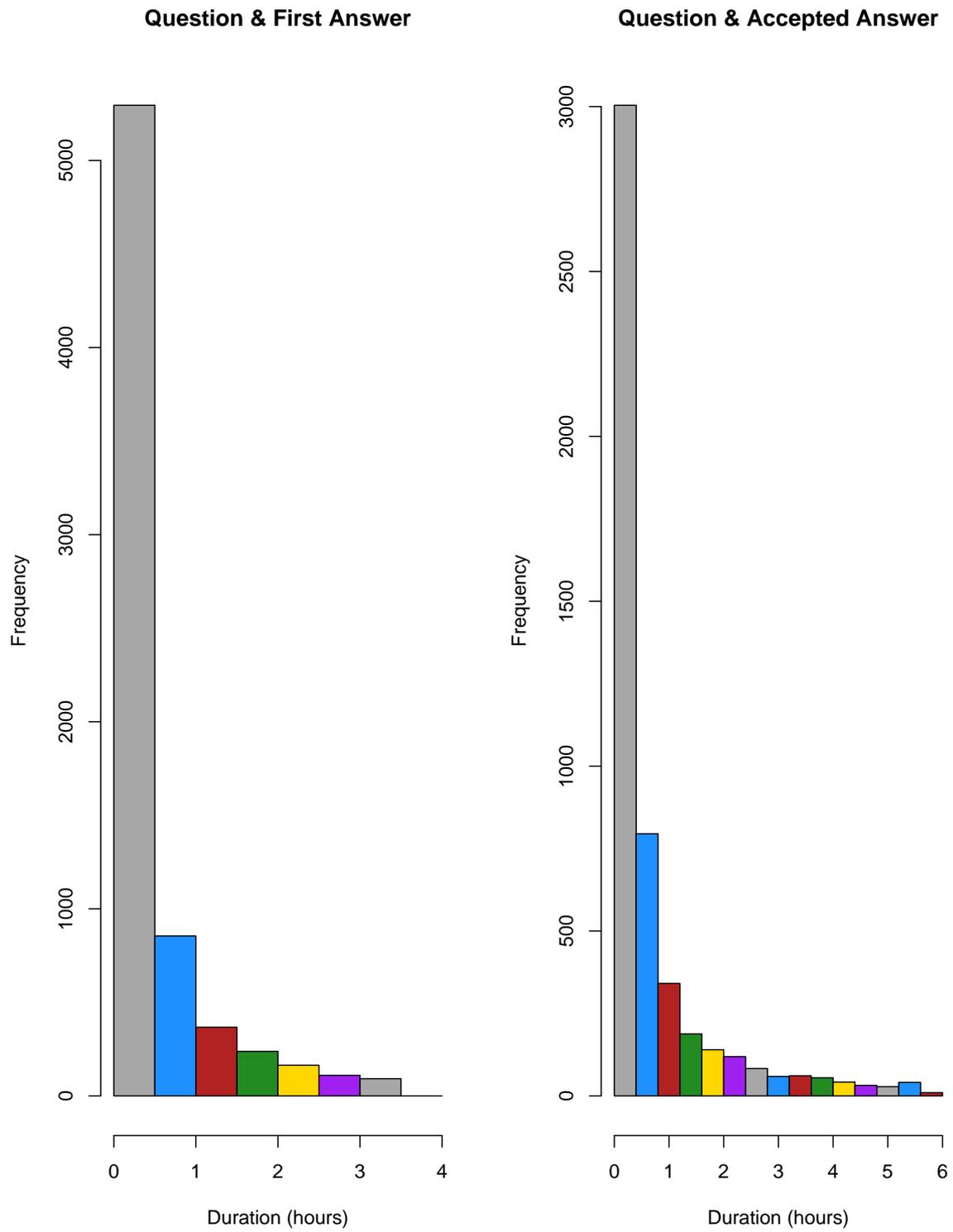}
 	\caption{Distribution of time (in hours) from when a question is asked until the reception of an answer.}
 	\label{Figure:chart_timeduration}
\end{figure}

Next, we look at the time duration between a developer asking a question and receiving a response. In this analysis, we look at how long it takes for a question to receive a response (regardless of it being an accepted answer) and how long it takes to receive an accepted answer. When examining the time duration between question-accepted answer pairs, we consider all refactoring posts in our dataset and do not restrict our analysis to pairs created in the same year. For completeness, we present the median values with and without outliers (removed via the Tukey Fences approach \cite{Jones2018Probability}). The median time between a question and its first answer is 0.27 hours with outliers and 0.19 hours without outliers. The median time between a question and an accepted answer is 0.41 hours with outliers and 0.27 hours without outliers. Additionally, in Figure \ref{Figure:chart_timeduration}, we also show histograms of time duration values (without outliers)\footnote{When reading/comparing these two histograms, it should be noted that the `Frequency' scale for the two charts differs.}. This chart shows that most responses to refactoring questions occur within the first hour of the developer asking the question.

%\begin{figure}
% 	\centering
% 	\includegraphics[width=1.0\linewidth]{chart_yearlybreakdown.pdf}
% 	\caption{Yearly breakdown of refactoring posts.}
% 	\label{Figure:chart_yearlybreakdown}
%\end{figure}

%Figure \ref{Figure:chart_yearlybreakdown} shows a yearly breakdown of refactoring related questions and their associated accepted answer posts. It should be noted that Stack Overflow was launched in September 2008, and our dataset contains posts up to March 2020, hence the relatively low number of posts for these two years when compared to the rest. Excluding 2008 and 2020, on average, each year contains approximately 833 questions, 538 accepted answers, and 1,226 other (i.e., non-accepted) answers. When looking at how long it takes for a question to receive a response, we observe it takes, on average, 
%27.78 minutes \ali{maybe a beanplot (or boxplot) would show the overall dataset} for the question to receive an answer (accepted or non-accepted), while the average time taken to receive an accepted answer is 43.19 minutes. \mohamed{How come the number of accepted answers is higher than the number of questions in 2010? How come non accepted answers is lower than accepted answers? Also why comparing apples with oranges? I was expecting to see questions without accepted answers vs. questions with accepted answers.}

\begin{figure}
 	\centering
 	\includegraphics[ trim=0cm 1.5cm 2cm 1.3cm, width=1\linewidth]{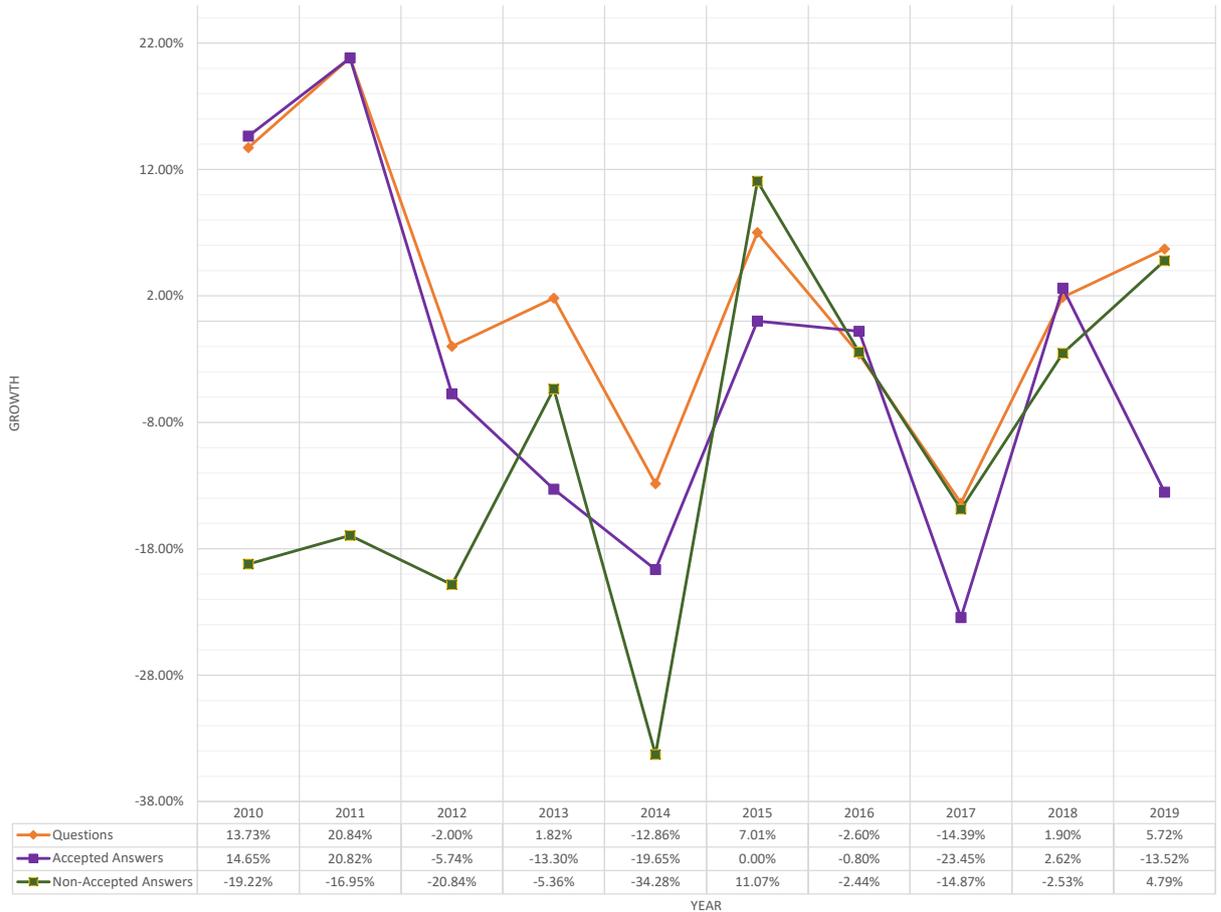}
 	\caption{Year-over-year growth of refactoring posts.}
 	\label{Figure:chart_yoyGrowth_posts}
\end{figure}

Looking at the year-over-year (YOY) growth around refactoring discussions (refer to Figure \ref{Figure:chart_yoyGrowth_posts}), we see that questions and accepted answers share a similar pattern between the YOY growth for questions and accepted answers. While the number of questions and accepted answers have increased, the volume by which they increased has been falling. Furthermore, to measure the extent of the relationship between these two variables, we conducted a Pearson correlation test \cite{Taeger2014Statistical}. This particular test is also known as a parametric correlation test as it depends on a normal distribution of the data, which we confirmed via a Shapiro-Wilk normality test \cite{Taeger2014Statistical}. The Pearson correlation test yielded a statistically significant (i.e., $p$-value $<$ 0.05) correlation coefficient of 0.872, equating to a strong correlation.

Finally, from the two charts, we observe a phenomenon where the number of questions forms a wave-like pattern throughout the years. Hence, while we see a dip in questions in 2017, there has been a gradual uptick in the subsequent years. This is an interesting phenomenon and would require further research to explain this pattern.

%\begin{tcolorbox}
%\textbf{Summary.}
%While it can be seen that developers continually seek advice and help from the community on refactoring related problems, the year-over-year growth has seen a decline since the initial years of Stack Overflow. However, almost 64\% of the questions have an accepted answer, showing that most developers are indeed finding solutions for their problems. Additionally, most of the accepted answers are received, on average, within an hour of posting the question. Finally, we observe a statistically significant correlation between the yearly volume of questions and accepted answers.
%\end{tcolorbox}

\subsubsection{\RQAB}
\subsubsection*{\textit{Motivation \& Approach:}}
After looking at the general trend of refactoring questions, this sub-RQ looks at the involvement of the Stack Overflow community in refactoring discussions. We investigate if a selected set of community members is responsible for asking and answering refactoring questions. To answer this sub-RQ, we capture the unique members posting questions and answers. More specifically, we utilize the \texttt{OwnerUserId} field to identify the creator of a post (i.e., question and answer).

\subsubsection*{\textit{Findings:}}
In the first part of our analysis, we look at the volume of unique users posting questions and answers. Looking at questions, we observe that 7,795 distinct users are responsible for creating all 9,489 refactoring-related questions. When it comes to answers, we found that 4,610 distinct users are associated with accepted answers, while 10,566 distinct users are associated with non-accepted answers.

Next, as shown in Table \ref{Table:DistinctUsers}, we look at the distribution of these unique users for each type of post. From the whole set of distinct users posting questions, most ($\approx$ 82.64\%) of the users post only questions. A similar pattern emerges when we look at the distinct set of users posting non-accepted answers; most ($\approx$ 79.65\%) of these users only post non-accepted answers. However, looking at the set of distinct users posting accepted answers, we see more of an even distribution. 

\begin{table}
\centering
\caption{Distribution of distinct user types.}
\label{Table:DistinctUsers}
\begin{tabular}{@{}lrr@{}}
\toprule
\multicolumn{1}{c}{\textbf{Item}} & \multicolumn{1}{c}{\textbf{Count}} & \multicolumn{1}{c}{\textbf{Percentage}} \\ \midrule
\multicolumn{3}{c}{\textit{Questions - total distinct user: 7,795}} \\
Users who post only questions & 6,442 & 82.64\% \\
Users who post questions \& accepted answers & 553 & 7.09\% \\
Users who post questions \& non-accepted answers & 800 & 10.26\% \\ \midrule
\multicolumn{3}{c}{\textit{Accepted Answers - total distinct user: 4,610}} \\
Users who post only accepted answers & 2,707 & 58.72\% \\
Users who post accepted answers \& questions & 553 & 12.00\% \\
Users who post accepted \& non-accepted answers & 1,350 & 29.28\% \\ \midrule
\multicolumn{3}{c}{\textit{Non-Accepted Answers - total distinct user: 10,566}} \\
Users who post only non-accepted answers & 8,416 & 79.65\% \\
Users who post non-accepted answers \& questions & 800 & 7.57\% \\
Users who post non-accepted \& accepted answers & 1,350 & 12.78\% \\ \bottomrule
\end{tabular}
\end{table}

In Stack Overflow, it is possible for the user who asks a question to answer their question and even mark their answer as an accepted answer. Therefore, we next look at such instances for refactoring posts. When it comes to accepted answers, we observe that 6.50\% (or 389) of question-accepted answer pairs have the same user creating the question and accepted answer. For non-accepted answer-question pairs, approximately 2.30\% (or 322) instances have the question creator also posting an answer.

\begin{table}
\centering
\caption{Frequency distribution of the number of posts created by a user.}
\label{Table:FrequencyDistribution}
\begin{tabular}{@{}lrr@{}}
\toprule
\multicolumn{1}{c}{\textbf{\begin{tabular}[c]{@{}c@{}}Number of posts \\ created by a user\end{tabular}}} & \multicolumn{1}{c}{\textbf{Count}} & \multicolumn{1}{c}{\textbf{Percentage}} \\ \midrule
\multicolumn{3}{c}{\textit{\begin{tabular}[c]{@{}c@{}}Questions - total distinct user: 7,795\end{tabular}}} \\
1 & 6,913 & 88.69\% \\
2 & 599 & 7.68\% \\
3 & 149 & 1.91\% \\ 
4 & 65 & 0.83\% \\ 
6 & 20 & 0.26\% \\ 
\textit{Others} & 49 & 0.63\% \\ \midrule
\multicolumn{3}{c}{\textit{\begin{tabular}[c]{@{}c@{}}Accepted Answers - total distinct user: 4,610)\end{tabular}}} \\
1 & 3,948 & 85.64\% \\
2 & 425 & 9.22\% \\
3 & 98 & 2.13\% \\
4 & 53 & 1.15\% \\
5 & 27 & 0.29\% \\ 
\textit{Others} & 59 & 1.28\%  \\ \midrule
\multicolumn{3}{c}{\textit{\begin{tabular}[c]{@{}c@{}}Non-Accepted Answers - total distinct user: 10,566)\end{tabular}}} \\
1 & 8,856 & 83.82\% \\
2 & 1,059 & 10.02\% \\
3 & 316 & 2.99\% \\ 
4 & 129 & 1.22\% \\ 
5 & 58 & 0.55\% \\ 
\textit{Others} & 148 & 1.40\%  \\ \bottomrule
\end{tabular}
\end{table}

Our final analysis looks at the volume of posts that a user creates. We observe that the majority of distinct users asking questions would ask, at most, only one question ($\approx$ 88.69\%). We observe similar patterns for accepted and non-accepted answers. In these two instances, approximately 85.64\% and 83.82\% of the distinct users create a single accepted and non-accepted answer post, respectively. Table \ref{Table:FrequencyDistribution} shows the top five distributions for each type of post.
%\begin{tcolorbox}
%\textbf{Summary.} We observe a segregation between community members that post and respond to questions. Members who mostly ask questions tend to stick with only asking questions. Furthermore, a community member posts, on average, a single refactoring question or answer to the site.
%\end{tcolorbox}

\subsubsection{\RQAC}
\subsubsection*{\textit{Motivation \& Approach:}} Tags are an essential part of Stack Overflow posts; they enable developers to categorize their questions and also help experts quickly access the questions that they specialize in. In this sub-RQ, we explore the types of tags that developers associate with refactoring questions. Our analysis of tags will help determine the concepts and technologies associated with refactoring challenges developers face. These findings provide a high-level overview, which we elaborate on in the subsequent RQs. To obtain the data for this sub-RQ, we extract all tags from all refactoring posts\footnote{In Stack Overflow, tags can only be associated with a question post.}.

\subsubsection*{\textit{Findings:}}
In total, our dataset contains 3,053 distinct tags. Not surprisingly, the most frequently occurring tag in our dataset is `refactoring'. We observe 6,808 ($\sim$21.89\%) questions containing the `refactoring' tag. Since this study is on refactoring, going forward, our analysis of tags will exclude counting the `refactoring' tag and instead focus on the other tags added by developers to refactoring posts to understand the area of refactoring better.

The top five tags were all related to programming languages (or web frameworks)-- Java (1,529 or 7.58\% instances), C\# (1,512 or 7.50\% instances), JavaScript (946 or 4.69\% instances), Ruby on Rails (591  or 2.93\% instances), and Ruby (569 or 2.82\% instances). These top five tags accounted for 27.82\% of the tags in the dataset. Additionally, most of the frequently occurring programming language tags in our dataset also appear in the list of top programming languages from 2016 to 2018 \cite{TopProgramming2016,TopProgramming2017,TopProgramming2018,TopProgramming2019}.

\begin{figure}
 	\centering
 	\includegraphics[clip, trim=0cm 1.5cm 2cm 1.3cm, width=1\linewidth]{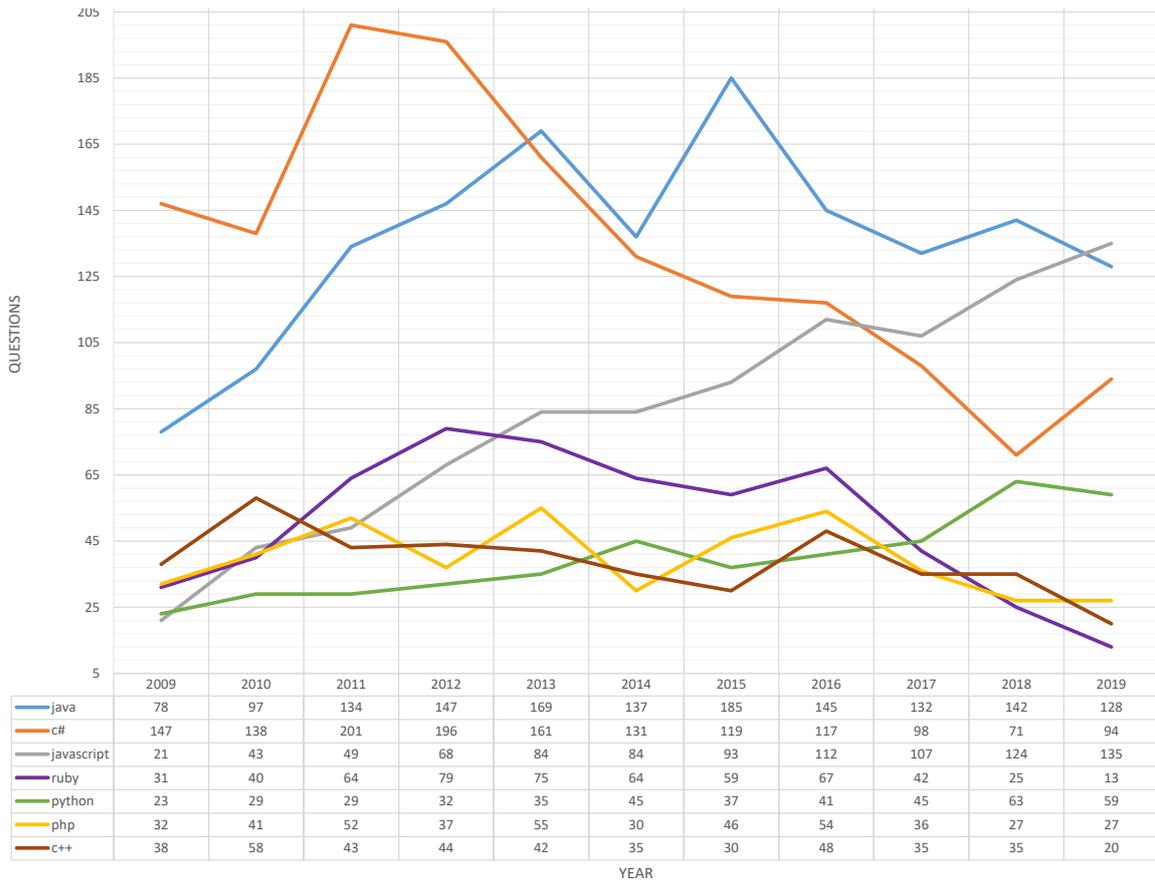}
 	\caption{Yearly growth of the popular tags associated with refactoring posts.}
 	\label{Figure:chart_yoyGrowth_tags}
\end{figure}

Next, as part of our analysis, we look at the yearly growth of the six most popular programming languages refactoring tags in the dataset. When reviewing the tags for each year,  we observe that these tags were part of a frequently occurring set of tags each year.  In Figure \ref{Figure:chart_yoyGrowth_tags}, we plot the volume by which questions associated with each tag either grow or shrink year-by-year (we ignore the years 2008 and 2020 as data is not available for the entire year). From this graph, we observe questions tagged with C\# show a steep decline from 2012, while at the same time, we see the volume of Java tagged posts being more-or-less constant. We also observe a constant increase in JavaScript tagged posts throughout the years; similarly, though relatively low in volume compared to C\# and Java posts, we also observe a steady increase in Python tagged posts. Furthermore, our findings on the rise of dynamically typed languages and the fall of statically typed languages are also in alignment with the Popularity of Programming Language Index \cite{PYPL}.

\begin{figure}
 	\centering
 	\includegraphics[clip, trim=0cm 2.5cm 0cm 2.5cm, width=.7\linewidth]{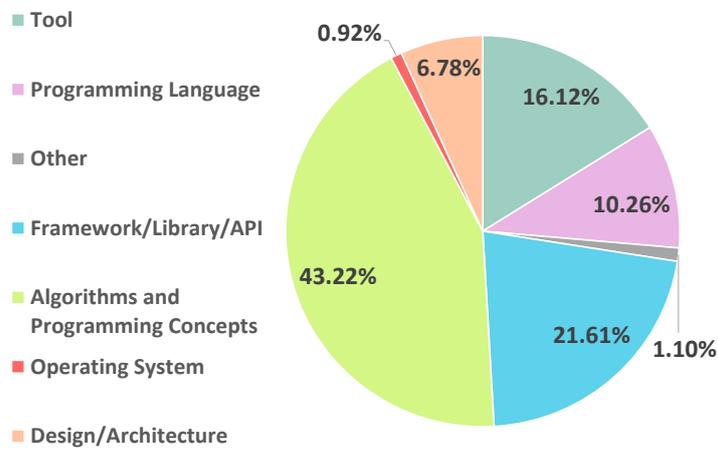}
 	\caption{Breakdown of the volume of instances for each tag category.}
 	\label{Figure:chart_tagcategory}
\end{figure}

To determine the different categories these tags fall under, we manually reviewed a statistically significant sample of tags. To this extent, two authors reviewed 547 of the frequently occurring tags; this sample corresponds to a confidence level of 99\% and a confidence interval of 5\%. When reviewing the utilized tags, each author made notes for each tag. Comparing notes, the authors discussed and settled on the finalized set of annotation categories. The finalized set consists of seven categories-- Tools, Programming Languages, Framework/Library/API, Algorithms and Programming Concepts, Design/Architecture, Operating Systems, and Other. Presented in Figure \ref{Figure:chart_tagcategory}, is a breakdown of the volume of instances for each tag category.

The majority of the tags (236 instances or 43.22\%) are related to general algorithms and programming concepts, which include code quality (e.g., `code-cleanup'), data types (e.g., `arrays'), and programming concepts (e.g., `recursion') among others. The next most popular category is frameworks/libraries/APIs (118 instances or 21.61\%). Some of the tags under this category include javascript-based frameworks and libraries (e.g., `reactjs'), web frameworks (e.g., `django'), non-web libraries such as testing (e.g. , `junit'), object-relational mapping (e.g., `hibernate'), and others (e.g., `pandas'). Tags falling under the tools category (88 instances or 16.12\%) include questions around IDEs (e.g., `eclipse'), database (e.g., `mysql'), and plugins (e.g., `resharper'), among others. The majority of the tags under programming languages (56 instances or 10.26\%) include those that support a multi-paradigm model such as `java', `c\#', and `python'; the next most common paradigm is declarative (e.g., `sql' and `html'). The design/architecture category includes tags related to patterns (e.g., `model-view-controller'). The popular tags in the operating system category are mobile-based (i.e., `android' and `ios'). Finally, the tag `iphone' is the most popular tag in the other category, containing six instances or 1.10\%. %\ali{I suggest to add a figure with "word cloud" if possible to visually show the different tags (easier to understand for the reader/reviewer)}

%Our analysis of tags helps to show the reason behind refactoring questions. However, these findings are at a very high-level since the number of available tags is, more-or-less, fixed. Hence, to gain more detailed insight, an analysis of the textual content of the post's body is required. In the following RQ, we examine the common terms in posts to achieve this.

%\begin{tcolorbox}
%\textbf{Summary.} In this RQ, we observe that the majority of refactoring questions are related to programming languages, more specifically Java and C\#. However, even though most existing posts are related to these static typed languages, we see a rise in questions around dynamic typed languages such as JavaScript and Python. Furthermore, we observe that the frequently occurring tags in our dataset align with research on programming languages popular among developers. Finally, we can group the 3,053 distinct tags in our dataset into five categories-- Tools, Programming Languages, Framework/Library/API, Software Engineering Concepts, Operating Systems, and Other. From this RQ, we envision IDE vendors and educators focus on empowering developers with the necessary skills and tools in refactoring code written in some of the most popular dynamically typed programming languages (e.g., Python and JavaScript) as developers are currently facing challenges in refactorings systems implemented with such programming languages.
%\end{tcolorbox}

\begin{tcolorbox}
\textbf{Summary for RQ$_1$.} 
Stack Overflow is a popular venue for developers to receive solutions for their refactoring questions, usually receiving a response in a short timeframe. Furthermore, while most questions involve statically typed languages, specifically Java and C\#, we see a rise in questions around dynamically typed languages such as JavaScript and Python. This finding shows a need for refactoring support from researchers and tool developers for dynamically typed languages and their unique issues. Finally, our tags analysis shows that most questions are around algorithm and programming concepts, followed by questions around specific frameworks/libraries.
% With almost 64\% of refactoring questions having an accepted answer, Stack Overflow is a popular venue for developers to seek advice and help from the community on refactoring related problems. It is also encouraging to observe that developers receive a response to their question within an hour from asking the question, on average. We also observe segregation between community members that post and respond to questions. Members who mostly ask questions tend to stick with only asking questions. Furthermore, while most questions involve statically typed languages, specifically Java and C\#, we see a rise in questions around dynamically typed languages such as JavaScript and Python. Furthermore, we observe that the frequently occurring tags in our dataset align with research on programming languages popular among developers. Finally, we can group the 3,053 distinct tags in our dataset into seven categories-- Tools, Programming Languages, Framework/Library/API, Design/Architecture, Algorithms and Programming Concepts, Operating Systems, and Other.
\end{tcolorbox}

\subsection{\textbf{\RQB}}
As a question and answer site, the fundamental purpose of Stack Overflow is to facilitate developers to utilize natural language to either describe the problem they require help with or provide advice/solutions to these problems. Prior work has shown that the language used to describe refactorings can vary heavily \cite{AlOmar2019IWoR,alomar2020toward}. For instance, the word `refactoring' is not always used (e.g., clean-up is a common alternative). Furthermore, the context surrounding the refactoring being applied can affect the particular alternative phrase that a developer chooses to use instead of the term `refactoring'. Thus, in RQ2, we explore the language used by developers to discuss refactorings through questions and answers. The outcomes from this question are meant to support a more robust analysis of refactoring rationale by revealing how developers discuss refactorings and what terminology frequently correlates in refactoring-related discussion topics. This RQ is composed of three sub-RQs. 
In  RQ$_{2.1}$, we examine the set of frequently occurring keywords in refactoring posts. In the second part of the research question, we utilize a refactoring taxonomy, known as Self-Affirmed Refactoring (SAR) \cite{AlOmar2019IWoR}, to define refactoring opportunities. This taxonomy contains internal structural metrics (coupling, complexity, etc.), external quality attributes (code comprehension, readability, reusability, etc.), and code smells (duplicate code, long methods, etc.). Existing studies have shown the existence of posts discussing the removal of code smells \cite{Tahir2018EASE,tahir2020large}, but little is known about how developers discuss the other types. Therefore, In RQ$_{2.2}$, we investigate the extent to which this taxonomy is triggering discussions around their refactoring. %developers explicitly document self-affirmed refactoring messages in their questions \ali{maybe also one sentence to say what does self-affirmed means (this is indicated once in the related work section, but people might not read/get it}.
Finally, in RQ$_{2.3}$, we perform a deep dive into the questions asked by developers by grouping related questions.

\subsubsection{\RQBA}
\label{SubSubSection:RQBA}
\subsubsection*{\textit{Motivation \& Approach:}}
The prior research questions show that developers do indeed encounter real-world challenges when refactoring their systems, and, at a high-level, our analysis of tags also indicates areas associated with refactoring challenges. However, since developers need to select the tags from a predefined list, we are limited in understanding more specific refactoring areas. Hence, in this sub-RQ, we extract the top keywords as bigrams from question posts. % and answer posts \eman{For this part, we consider both question and answer posts, whearas in the discussion of SAR, we consider only question posts.}. 
Bigrams correspond to a sequence of two adjacent words in a sentence. We did look at trigrams, but we could not locate sets of common terms. Unlike unigrams, bigrams provide a certain level of context for terms, which helps our analysis by reducing the chance of making false presumptions. Additionally, we also examine the existence of specific refactoring terms. These terms correspond to refactoring operations defined in Fowler's catalog of refactoring operations (e.g., extract method, move method, rename attribute, etc.) \cite{Fowler2018refactoring}. %By analyzing the natural language in the posts, we obtain more specific refactoring related terminology utilized by developers.

\subsubsection*{\textit{Findings:}}
In Table \ref{Table:Common_Bigrams}, we present the top ten frequently occurring bigrams in question posts. % and answer posts. 
This table shows that the IDE `visual studio' plays an important part in refactoring discussions. Since this particular IDE can support multiple types of programming languages, it has become popular among developers in implementing systems \cite{PYPL_IDE}. The mention of the IDE falls into two categories-- (1) developers asking for assistance in performing a refactoring operation using the IDE, or (2) developers mentioning the IDE when providing context around the refactoring challenge they encounter. Furthermore, the bigram `refactoring tool' also emphasizes the importance and reliance of tools and IDEs in refactoring activities. We also observe discussions around the refactoring of test suites showing that interest in refactoring source code among developers is not just limited to production code. The bigram `legacy code' highlights a common reason why developers request support with refactoring. In the subsequent sub-RQ's (RQ$_{2.2}$ and RQ$_{2.3}$), through qualitative analysis, we further contextualize most of these bigrams and provide exemplar posts from where they were extracted.

Since the names of refactoring operations do not occur in the top ten frequently occurring bigrams list, we investigate if developers use these terms in their questions. These are the terms defined in Fowler's catalog of refactoring operations \cite{Fowler2018refactoring}. We observe that the bigram `extract method' frequently occurs in question posts. However, we should also note that we also encounter instances where developers talk about extracting other types of identifiers, such as classes, interfaces, and variables. For move operations, once more, we observe that developers frequently mention the `move method' operation. However, we also observe instances of moving classes and files. We also observe the terms `common' and `code' associated with the refactoring operation terms `extract' and `move', showing that developers discuss refactoring operations without using the standard terms defined in the catalog or refactoring operations. Looking at inline operations, once more, methods were mentioned frequently for inlining. However, looking at rename operations, we observe more occurrences of files, classes, packages, and variables than methods. Finally, push-up and pull-down operations did not yield anything significant (similar to findings by Danilo et al. \cite{Danilo2016FSE}).

This sub-RQ highlights common bigrams in refactoring posts to understand why developers utilize Stack Overflow for refactoring discussions. These bigrams are phrases associated with refactoring operations, showing that developers are aware of refactoring concepts. Going one step forward, in the next sub-RQ, we explore the use of other known software engineering refactoring phrases.

\begin{table}
\centering
\caption{Top ten frequently occurring bigrams for question posts.}
\label{Table:Common_Bigrams}
\begin{tabular}{@{}lrr@{}}
\toprule
\multicolumn{1}{c}{\textbf{Bigram}} & \multicolumn{1}{c}{\textbf{Count}} & \multicolumn{1}{c}{\textbf{Percentage}} \\ \midrule
visual studio & 348 & 1.57\% \\
unit test & 253 & 1.14\% \\
base class & 164 & 0.74\% \\
create new & 160 & 0.72\% \\
refactoring tool & 155 & 0.70\% \\
switch statement & 146 & 0.66\% \\
business logic & 133 & 0.60\% \\
design pattern & 132 & 0.59\% \\
legacy code & 115 & 0.52\% \\
add new & 110 & 0.50\% \\
\textit{Others} & 20,484 & 92.27\% \\ \bottomrule
\end{tabular}
\end{table}

\subsubsection{\RQBB}
\label{SubSubSection:RQBB}
\subsubsection*{\textit{Motivation \& Approach:}}
The concept of Self-Affirmed Refactoring (SAR) introduced by AlOmar et al.  \cite{AlOmar2019IWoR,alomar2020toward} explores how developers document their refactoring activities %, in order to gain more insights about them, 
 such as the intent behind the refactoring type of operations performed. 
%i.e., developers explicitly document refactorings intentionally introduced during a code change. 
In their work, the authors identify recurring patterns in SAR commit messages and define three SAR categories, (1) internal quality attributes, (2) external quality attributes, and (3) code smells. In our work,  we use these SAR terminology patterns in addition to other keywords related to code smells reported in literature \cite{Fowler2018refactoring,Tahir2018EASE} that belong to each of the three SAR categories as indicators of refactoring activity-related discussions. %In other words, if a pattern exists in a question post, it is then considered as a SAR post. 
In other words, we string match SAR patterns in question posts to see the extent to which they contribute to challenges developers face when they refactor. We also extract the bigrams from question posts to better understand the context of SAR patterns and help us in our analysis. We are particularly interested in extracting the intent behind the refactoring in questions to capture what typically triggers developers to refactor their code.

\subsubsection*{\textit{Findings:}}
Table \ref{Table:SAR_Patterns} depicts the list of SAR patterns, ranked based on their frequency, we identify in questions. We observe that developers frequently mention key internal quality attributes (such as inheritance, cohesion, etc.) and a wide range of external quality attributes (such as readability and performance), and a variety of code smells that might impact code quality. Upon closer inspection of the generic refactoring patterns, we notice that developers use a variety of patterns to discuss refactorings such as `clean up' or `redesign', although `refactor' is the most used keyword (29.09\%). Additionally, these patterns are mainly linked to code elements at different levels of granularity (e.g., `add an attribute', `create an interface', `refactor the method'). Further, we observe that developers mention the motivation driving refactorings that are not restricted only to fixing code smells, as in the original definition of refactoring in Fowler's book \cite{Fowler2018refactoring}. %which are aligned with the finding from the previous studies \cite{Danilo2016FSE,alomar2020howwe,Kim2014TSE}. 
We also observe that developers tend to report the executed refactoring operations using keywords such as `extract' or `rename'.

To improve the internal design, the optimization of dependency seems to be the dominant focus that is consistently mentioned (41.86\%). It is apparent from some of the posts that developers intend to introduce best practices (e.g., the use of object-oriented design principles, the application of inheritance, polymorphism, and optimization of software quality metrics to reduce code complexity). Further, developers refactor the code to improve the dominant modularization driving forces (i.e., cohesion and coupling) to maximize intra-class connectivity and minimize inter-class connectivity. 

Concerning external quality attribute-related questions, we observe the mention of refactorings to enhance nonfunctional attributes. Terms such as `readability', `efficiency', and `performance' represent the developers' main focus, with 16.37\%, 13.85\%, and 11.52\%, respectively. Although multiple studies \cite{pantiuchina2018improving,fakhoury2019improving,alrubaye2020does} have been analyzing code comprehension and using metrics to measure readability, there is no mention of these readability tools/models (i.e., \cite{dorn2012general,scalabrino2018comprehensive,buse2009learning,posnett2011simpler}) in the questions. For instance, developers refactor the code to improve its reusability. More recently, AlOmar et al. \cite{alomar2020developers} show that the number of methods significantly increases when developers refactor the code to improve reusability. Also, developers make changes such as extracting methods to improve testability as they test parts of the code separately. Developers also extract methods to improve code readability. 

Finally, for code smell-focused refactoring questions, we observe that duplicate code represents the most popular anti-pattern that developers intend to refactor (28.75\%). While there are also various tools for detecting and potentially refactoring code clones \cite{kamiya2002ccfinder,roy2009comparison,mazinanian2016jdeodorant}, we could not locate any reference to them. Also, developers perform refactorings to eliminate specific code smells (e.g., spaghetti code, long method, feature envy, etc.) that are known to deteriorate the quality of the source code. For example, based on our manual analysis, we observe that developers discuss performing `Extract Method' refactoring to remove a code smell, which corresponds to a long method (i.e., a bad smell). Developers also indicate the generic pattern `code smell' in addition to the specific name of the code smell under correction.

This sub-RQ shows that SAR patterns documented in commit messages are also utilized by developers when crafting questions on Stack Overflow. While these phrases/terms are indicative of refactoring activities, they are specific to SAR patterns and, at a high-level, do not provide context as to where or what is associated with the terms. Hence, in the following sub-RQ, we group all related terms (i.e., SAR and non-SAR pattern terms) to determine the primary categories that cause developers to seek assistance when refactoring. We perform this grouping on the entire set of questions on our dataset via the use of an unsupervised machine learning technique to determine the refactoring topics associated with refactoring questions.

%Generally, we found that these SAR patterns are documented by Stack Overflow developers in practice to indicate refactoring-related discussions (similar to what developers documented in their commit messages \cite{AlOmar2019IWoR,alomar2020toward,alomar2020howwe}).

\begin{table}
\centering
\caption{Frequency of Self-Affirmed Refactoring (SAR) patterns in questions posts.} %\eman{I added this table to reflect SAR patterns. I am suggesting to merge the discussion about them with what is written about Table 5}}
\label{Table:SAR_Patterns}
\begin{tabular}{@{}lr|lll@{}}
\toprule
\multicolumn{2}{c|}{\textbf{Generic SAR}} & \multicolumn{3}{c}{\textbf{Specific SAR}} \\
\multicolumn{1}{c}{\textbf{Keyword}} &  \multicolumn{1}{c|}{\textbf{Percentage}} & \multicolumn{1}{c}{\textbf{Internal QA}} & \multicolumn{1}{c}{\textbf{External QA}} & \multicolumn{1}{c}{\textbf{Code Smell}} \\ \midrule
refactor* &  29.09\% & dependency (41.86\%) & readability (16.37\%) & duplicate code (28.75\%) \\
chang* &  10.02\% & complexity (14.06\%) & efficiency (13.85\%) & switch statement (19.23\%) \\
creat* &  9.12\% & inheritance (13.27\%) & performance (11.52\%) & code smell (12.27\%) \\
add*  &  8.30\% & polymorphism (8.05\%) & configurability (6.78\%) & case statement (6.95\%) \\
mov* &  7.05\% & coupling (7.74\%) & productivity (6.08\%) & redundancy (4.76\%) \\
clean* &  3.64\% & abstraction (7.26\%)& effectiveness (5.67\%) & spaghetti code (4.02\%) \\
renam* &  3.61\% & composition (4.73\%) & maintainability (5.55\%) & anti-pattern (4.02\%) \\
remov* &  3.27\% & encapsulation (2.05\%) & usability(4.73\%) & long method (3.84\%) \\
replac* &  3.07\% & cohesion (0.94\%) & testability (4.44\%) & dead code (3.66\%) \\
extract*  & 2.71\% &  & compatibility (3.85\%) & large class (2.38\%)  \\
fix* &  2.40\% &  & reusability (3.33\%) & god class (1.64\%) \\
improv* &  2.12\% &  & modularity (2.22\%) & technical debt (1.46\%) \\
modif* &  1.83\% & & flexibility (2.16\%) & bad smell (1.46\%) \\
simplif* &  1.59\% &  & accessibility (1.98\%) & data class (1.46\%) \\
rewrit* &  1.41\% &  & manageability (1.92\%) & long parameter list (0.91\%) \\
reduc* &  1.38\% &  & reliability (1.57\%) & complex class (0.54\%) \\
split* &  1.26\% &  & simplicity (1.40\%) & middle man (0.54\%) \\
extend* &  1.05\% &  & extensibility (1.16\%) & code clone (0.54\%) \\
introduc* &  1.03\% &  & accuracy (1.05\%) & inappropriate intimacy (0.36\%) \\
merg* & 0.74\% &  & stability (0.99\%) & feature envy (0.36\%) \\
pull* &  0.71\% &  & understadability (0.93\%) & anti pattern (0.36\%) \\
get rid of* & 0.66\% &  & robustness (0.70\%) & brain class (0.18\%) \\
migrat* &  0.66\% &  & scalability (0.64\%) & message chain (0.18\% )\\
organiz* & 0.64\% &  & correctness (0.40\%) &  \\
inlin* &  0.49\% &  & modifiability (0.23\%) & \\
push* &  0.41\% &  & adaptability (0.11\%) & \\
tidy* &  0.19\% & & reproducibility (0.11\%)& \\
restructur* &  0.19\% &  & repeatability (0.05\%) & \\
customiz* &  0.18\% &  & interoperability (0.05\%) &  \\
redesign* &  0.18\% & &  &  \\
aggrega* &  0.15\% &  & &  \\
reformat* &  0.13\% &  &  &  \\
reorganiz* & 0.13\% & &  &  \\
enhanc* &  0.11\% &  &  &  \\
decompos* &  0.09\% &  &  &  \\
rework* &  0.08\% &  &  &  \\
modulariz* & 0.04\% &  &  & \\
refin* &  0.03\% &  &  &  \\
polish* &  0.02\% & &  &  \\
re packag* & 0.005\% & &  &  \\ \bottomrule
\end{tabular}
\end{table}

\subsubsection{\RQBC}
\label{SubSubSection:RQBC}
\subsubsection*{\textit{Motivation \& Approach:}}
The prior sub-RQs looked at common and SAR pattern terms in refactoring posts. In this sub-RQ, we go a step forward by grouping related terms to identify and understand the different areas (or topics) that developers require assistance with and understand the motivation behind refactoring.

We tackle this research question from three fronts. First, we perform an n-gram analysis to identify the common phrases developers utilize when describing their problem/challenge. Next, we perform a topic modeling analysis to identify the key topics associated with refactoring related questions. Finally, we manually analyze a statistically significant sample of questions to gain more insight and context around the detected topics. Topic modeling is an unsupervised machine learning procedure that infers the topics (or thematic structure) discussed in large volumes of unlabeled and unstructured text documents \cite{Johnston2019Applied}. N-grams are sets of co-occurring words (or letters), within a given window, that are available in a textual document and are useful in understanding a word in its context \cite{Jurafsky2009Speech}.

Prior to our topic modeling and n-gram analysis, we perform a set of pre-processing activities on the textual data. Some of our key pre-processing activities included: expansion of word contractions (e.g., `I\textquotesingle m' $\rightarrow$ `I am'), removal of URLs, code blocks, stopwords, alphanumeric words, and punctuations, retaining only nouns, verbs, adjectives, and adverbs and lemmatization of words. We opted to use lemmatization over stemming, as the lemma of a word is a valid English word \cite{lane2019natural}. In addition to the default set of stopwords supplied by NLTK \cite{Bird2002NLTKTN}, we added our own set of custom stop words. To derive the set of custom stop words, we generated and manually analyzed the set of frequently occurring words in our corpus. Examples of custom stop words include `thanks', `question', `answer', etc. 
We utilize the Latent Dirichlet Allocation (LDA) \cite{Blei2003LDA} algorithm for our topic modeling analysis. Our use of LDA for the topic modeling analysis follows prior research based on Stack Overflow posts \cite{Openja2020ICSME,Wang2013SAC,Bandeira2019MSR,Barua2014ESE,Rosen2016ESE,Villanes2017SBES,Ahmed2018ESEM,Yang2016JCST,Bangash2019MSR,Alshangiti2019ESEM,Allamanis2013MSR,Bagherzadeh2019FSE,Abdellatif2020MSR} that have shown the effectiveness of LDA in similar contexts. Essentially, LDA builds a statistical model that groups related words together from a corpus of textual documents where each grouping of frequently co-occurring words represents a topic. As the topics are not labeled, subject matter experts are then needed to determine each topic's name based on the analysis of the list of words. A mandatory input for the LDA algorithm is the number of topics to be generated. A low value will result in high-level or general topics, while a high value will produce more detailed topics, some of which will be noise. Hence, to arrive at the optimal number of topics, we iteratively extracted topics from two to fifty in increments of one. Each LDA execution cycle (i.e., model creation) was subjected to ten passes and one hundred iterations. In each cycle, we extracted the topic coherence \cite{Roder2015Coherence}, perplexity score \cite{Blei2003LDA}, and topic visualization \cite{Sievert2014LDAvis} of the model. Finally, to determine the optimal number of topics for our LDA analysis, we relied on a combination of topic coherence, perplexity, visualization, and manual analysis. The complete set of coherence and perplexity values for each of the fifty models and an interactive visualization for the optimum model is available on our project website. Concerning our manual and visual analysis-- we look at the topics and terms generated in each execution cycle of the LDA algorithm to discover patterns in the topics such as similarities and overlapping of topics, topics that are consistent between each execution cycle, the prevalence of each topic, distribution and relevance of words by topics, etc. Finally, since the LDA process does not result in meaningful names for the topics it generates, we manually examined the list of generated terms to determine the appropriate topic names. For our manual analysis, we undertook a  collaborative approach-- we looked at the terms that represent each topic, came to an agreement on the name of the topics, and identified the topics generated by noisy terms. Additionally, we also looked at the terms that are unique to each topic and the terms shared among topics (including the overall frequency of the term).

%\begin{figure}
% 	\centering
% 	\includegraphics[clip, trim=2cm 1.5cm 2cm 1.3cm, width=1\linewidth]{chart_yearlytopic.pdf}
% 	\caption{Yearly growth for the LDA topics associated with refactoring posts.}
% 	\label{Figure:chart_yearlyGrowth_topics}
%\end{figure}

\subsubsection*{\textit{Findings:}} 
%\eman{It would also interesting if we can show refactoring topics evolution over time if we have the data.}\anthony{done}
From our LDA analysis, we observe that the most optimal model yields five topics, associated with Stack Overflow refactoring questions: \textit{Code Optimization}, \textit{Architecture and Design Patterns}, \textit{Unit Testing}, \textit{Tools and IDEs}, and \textit{Database}. To understand the distribution of each topic in the dataset, we assigned the most dominant topic to each question. Our results show that \textit{Code Optimization} is the most frequently occurring topic (at 43.72\% or 4,142 questions). We present, in Table \ref{Table:LADTopicsWords}, the distribution for each topic in the dataset. Additionally, the table also shows a partial set of words (unigrams and bigrams) associated with each topic.

While examining the words occurring in a topic helps in determining the name of the topic, it does not adequately help in determining the rationale for the topic. These topic-based words alone do not indicate the problems developers encounter or the advice they solicitor around refactoring. Hence, we perform a manual analysis of a stratified statistically significant set of questions. Using a confidence level of 95\% and an interval of 10\% for each topic, we constructed a sample size of 430 posts that were analyzed by two authors. In this process, two authors annotated the dataset with the rationale behind the topic. Next, the authors exchanged the annotated datasets for review. During the review, if the reviewer disagreed with a specific annotation, the instance was marked for discussion. Finally, the annotator and reviewer discussed and looked at resolving the identified conflicts.

%In Figure \ref{Figure:chart_yearlyGrowth_topics}, we show the yearly growth of each topic from 2009 to 2019.

For each of the detected topics, we describe our findings and summarize the key challenge(s) and the takeaways for relevant stakeholders. Additionally, to provide context around topics, we include representative examples to Stack Overflow posts in the form of quotes.

\begin{table*}
\centering
\caption{LDA topics with their frequency of occurrence and a partial set of their corresponding words.}
\label{Table:LADTopicsWords}
\begin{tabular}{lrrl}
\hline
\multicolumn{1}{c}{\multirow{2}{*}{\textbf{Topic}}} & \multicolumn{2}{c}{\textbf{Question}} & \multicolumn{1}{c}{\multirow{2}{*}{\textbf{\begin{tabular}[c]{@{}c@{}}Key Words\\ (unigrams \& bigrams)\end{tabular}}}} \\
\multicolumn{1}{c}{} & \multicolumn{1}{c}{\textbf{Count}} & \multicolumn{1}{c}{\textbf{Percentage}} & \multicolumn{1}{c}{} \\ \hline
\multirow{3}{*}{Code Optimization} & \multirow{3}{*}{4,142} & \multirow{3}{*}{43.72\%} & loop, array, function, variable, operator, parameter, lambda,\\ 
 &  &  & dictionary, repeated, change, switch\_statement, helper\_method,\\
 &  &  & lambda\_expression, avoid\_duplication, global\_variable \\ \hline
\multirow{3}{*}{Tools and IDEs} & \multirow{3}{*}{2,395} & \multirow{3}{*}{25.28\%} & rename, package, resharper, file, tool, folder, import, ide,\\
 &  &  &  error, change, visual\_studio, android\_studio, separate\_file,\\
 &  &  &  intellij\_idea, command\_line \\ \hline
\multirow{3}{*}{Architecture and Design Patterns} & \multirow{3}{*}{1,660} & \multirow{3}{*}{17.52\%} & class, method, object, interface, model, abstract, subclass, \\
 &  &  & inherit, singleton, factory, base\_class, design\_pattern, \\
 &  &  & business\_logic, derived\_class, view\_controller \\ \hline
\multirow{3}{*}{Unit Testing} & \multirow{3}{*}{1,086} & \multirow{3}{*}{11.46\%} & test, unit, mock, logical, tdd, testable, coupled, coverage, \\
 &  &  & junit, proxy, unit\_test, business\_logic, test\_case,\\
 &  &  & write\_test, add\_new \\ \hline
\multirow{3}{*}{Database} & \multirow{3}{*}{190} & \multirow{3}{*}{2.01\%} & table, column, row, field, join, mysql, schema, dto, subquery,\\
 &  &  & entity, stored\_procedure, primary\_key, sql\_query,\\
 &  &  & foreign\_key, sql\_server \\ \hline
\end{tabular}
\end{table*}
%%%%%%%%%%%%%%%%%%%%%%%%%%%%%%%%%%%%%%%%%%%%%%%
\subsubsubsection{Code Optimization} \label{section:lda_codeoptimization}
% One of the common reasons developers look for help with code optimization is related to analyzability. In other words, developers want to improve the readability of their code. For instance, developers look for help in either simplifying or replacing switch statements (e.g., Quote~\ref{Quote:CodeOptimization_01}), compacting logic through the reduction of lines of code (e.g., Quote~\ref{Quote:CodeOptimization_02}), and through the removal of duplicate code (e.g., Quote~\ref{Quote:CodeOptimization_03}).

% Additionally, we also observe that developers run into compilation and runtime issues when refactoring their code and turn to the community for help in resolving the issues. For example, in this question-- Quote~\ref{Quote:CodeOptimization_04}, the developer performs an extract method operation as a means to reduce code duplication but then encounters a runtime exception. However, exceptions are not the only problems developers encounter when they refactor their code. For instance, in this question-- Quote~\ref{Quote:CodeOptimization_05}, the developer simplifies a nested loop by using LINQ, but runs into issues trying to preserve the existing behavior.

Program comprehension is a crucial-enough concern for developers that they turn to the community for assistance with improving the analyzability or readability of their source code. To this end, a common challenge developers face is reducing lines of code. Our analysis of exemplar posts (e.g., Quote~\ref{Quote:CodeOptimization_}) shows that developers seek assistance with performing refactoring operations involving code extraction and advice on any patterns that they should follow. Specifically, some common challenges include simplifying or replacing switch statements and loops, compacting logic, and removing duplicate code. In most instances, developers are dealing with code that consists of a series of complicated logic conditions that require significant and careful refactoring so as to not result in a break of functionality. Within this topic, we also observe situations where developers reach out to the community for help with resolving runtime and behavioral issues they encounter after making readability improvements to their code. Thus, showing that improving program comprehension is not always straightforward-- it can be both time-consuming and error-prone.

Next, observing the frequent terms in this topic (refer to Table \ref{Table:LADTopicsWords}), we encounter terms related to readability, such as `repeated', `helper method,' and `avoid duplication'; further highlighting that developers seek assistance with improving code reusability as a means of improving overall program comprehension. More specifically, this improvement often leads to optimization of switch-case statements and eliminating duplicate code, and possibly even improve the efficiency and performance of the system. Finally, it is worth noting that even though developers seek help with refactoring their code, they rarely utilize established refactoring terminology (e.g., `Extract Method') when describing their problem; instead, they tend to be more colloquial in their description.

\begin{center}
\fbox{\parbox{\dimexpr\linewidth-2\fboxsep-2\fboxrule\relax}{\centering
\textbf{How can I refactor this Python code to make it more readable and compact?}

\begin{flushleft}
\textit{``I wrote a function to handle selecting and using items to regain the player's health in a text adventure. What would be the best way to shrink the following code? Any feedback or constructive criticism would be greatly appreciated.''}
\end{flushleft}
}}
\captionof{Quote}{Sample question for the code optimization topic highlighting the need for assistance with improving program comprehension
\cite{lda_codeoptimization02}.\label{Quote:CodeOptimization_}}
\end{center}

In summary, this topic shows that developers primarily seek assistance with simplifying code structures to improve readability, and reusability, specifically around reducing the complexity caused by lengthy switch-case statements, loops, and duplicate code. This challenge presents the research community with two important opportunities: 1) to conduct studies around the automatic detection and refactoring of lengthy conditional code blocks (such as switch-case statements), and 2) to understand and measure the influence of different code structures, patterns, and architectures on comprehension (i.e., what is the best way to refactor a large group of nested conditional statements?). This challenge is related to prior work, which shows that readability metrics are currently struggling to measure real readability \cite{fakhoury2019improving}. 

\subsubsubsection{Tools and IDEs} \label{section:lda_tools}
% Questions pertaining to activities that are not about code, but yet relate to programming fall into this topic. When looking at questions under this topic, we observe that developers frequently ask questions about using IDEs (or tools) to perform rename operations. It should be noted that questions around renaming are not related to source code identifiers. There are instances where developers seek advice on renaming database elements (e.g., Quote~\ref{Quote:Tool_01}), packages (e.g., Quote~\ref{Quote:Tool_02}), files (e.g., Quote~\ref{Quote:Tool_03}), and content within other non-source code files such as XML (e.g., Quote~\ref{Quote:Tool_04}).

% Other than renaming, developers also ask for help around performing batch operations such as move operations (e.g., Quote~\ref{Quote:Tool_05}), type changes (e.g., Quote~\ref{Quote:Tool_06}), duplicate checking/elimination (e.g., Quote~\ref{Quote:Tool_07}). Our observations also show that developers seek advice from the community for recommendations on tools/IDEs to perform certain refactoring tasks (e.g., Quote~\ref{Quote:Tool_08},\ref{Quote:Tool_09},\ref{Quote:Tool_10}). Finally, we also encounter questions where developers require assistance with configuring their tools (e.g., Quote~\ref{Quote:Tool_11},\ref{Quote:Tool_12}).  

This topic deals with questions about activities related to tool-based refactoring. Tools are a vital part of software development. They provide developers with the means to automate time-consume code changes, thereby improving developer productivity and potentially eliminating the injection of defects. Predominately, questions around this topic are related to renaming activities. This aligns with the previous topic where we observe developers seeking assistance to optimize their code to improve program comprehension. However, questions around tool-based renaming are not just related to a straightforward renaming of a source code identifier. Instead, developers seek assistance with renaming other software engineering artifacts such as packages, database elements, files, and content within other non-source code files such as XML or performing bulk/batch-based renaming operations (e.g., Quote~\ref{Quote:Tool_}). In most instances, the questions are around using the rename functionality of their IDE.

While most questions we observe are around renaming, we also encounter questions about performing other refactoring operations such as move operations, type changes, duplicate checking/elimination. Once again, these questions are specific to the developer's project and, in most cases, nontrivial. Additionally, developers seek advice for recommendations for tools/IDEs for refactoring specific programming languages or for configuring tools (such as disabling/enabling specific IDE refactoring features). In their study of the usability of refactoring tools, Eilertsen and Murphy \cite{Eilertsen2021SANER} highlight the need for tools to support developers in guiding tools in the execution of refactoring operations. This corroborates our findings, where most questions around tool usage are specific to a developer's source code. Furthermore, many of the standard refactoring operations available by the tool do not meet the developer's unique refactoring needs. Lastly, from Table \ref{Table:LADTopicsWords}, we observe that questions involve using popular IDEs, specifically Visual Studio, IntelliJ IDEA, and Android Studio.  %Similarly, reviewing the Self-Affirmed Refactoring patterns (refer to Table \ref{Table:SAR_Patterns}), we observe that tools play a vital part in assisting developers with refactoring activities like the detection and resolving of code/design smells and anti-patterns.

\begin{center}
\fbox{\parbox{\dimexpr\linewidth-2\fboxsep-2\fboxrule\relax}{\centering
\textbf{Rename/Refactor database elements - only scripts exists but not database}

\begin{flushleft}
\textit{``I have script files ready to create a database. I also have coding standards/conventions to be set against those scripts. Is there a best/Easy way to rename these (according to coding standards I have) such that when I rename a database object, other objects that reference the renamed object should automatically updated with the new name.''}
\end{flushleft}
}}
\captionof{Quote}{Sample question for the tools/IDEs topic showing developers seeking assistance using tools/IDEs to perform nontrivial rename operations
\cite{lda_tool01}.\label{Quote:Tool_}}
\end{center}

In summary, this topic shows that developers understand the benefits of using tools to automate refactoring activities. However, the developer's expectation goes beyond the general/standard features offered by the tools. The primary challenge is with seeking assistance with renaming other software engineering artifacts outside of identifier names. These are artifacts such as packages, database elements, files, and others. Based on this, vendors should enhance their IDE's renaming facility to support renaming other software elements/artifacts (e.g., database elements) related to renamed source code identifiers. Furthermore, the research community needs to investigate models and techniques that can learn from and adapt to a developer's codebase. Such as providing better recommendations and improved detection of opportunities to refactor artifacts outside of code. Especially those artifacts that have direct links to entities in the code that are touched by a refactoring.

% \textcolor{red}{\subsubsubsection*{\textit{\textbf{Primary Challenge}}}
% Developers understand the benefits of using tools to automate refactoring activities. However, the developer's expectation goes beyond the general/standard features offered by the tools. The primary challenge is with seeking assistance with renaming other software engineering artifacts outside of identifier names. These are artifacts such as packages, database elements, files, and others. }

% \textcolor{red}{\subsubsubsection*{\textit{\textbf{Key Takeaways}}}
% \underline{\textit{For Tool Vendors:}} While almost all IDEs provide renaming capabilities, vendors should enhance this facility to support renaming other software elements/artifacts (e.g., databases) related to renamed source code identifiers. Furthermore, vendors should collaborate with developers to implement the optimal user experience to promote tool usage.}

% \noindent\textcolor{red}{\underline{\textit{For Researchers:}} The research community needs to invest in building tools that can learn from and adapt to a developer's codebase and, hence, provide appropriate refactoring recommendations during implementation. Furthermore, there is a research opportunity to examine how to automatically synchronize refactoring changes between related artifacts (e.g., data access layer and database). Finally, there is a need for a comprehensive survey or online catalog of refactoring tools, including a feature-by-feature comparison..}

%%%%%%%%%%%%%%%%%%%%%%%%%%%%%%%%%%%%%%%%%%%%%%%
\subsubsubsection{Architecture and Design Patterns} \label{section:lda_design}

As part of their evolution, industrial systems typically undergo architecture refactoring to maintain their structural quality \cite{Samarthyam2016IWOR}. A review of a sample set of posts on this topic shows that reusability is a crucial motivator for developers to refactor their systems. Developers primarily seek assistance with reducing the complexity of their systems, which in most accounts are legacy systems. Typically such systems have been online for a considerable period and undergone many updates by multiple developers. To this extent, the questions revolve around asking for assistance with adhering to design principles such as SOLID, DRY, SRP, and KISS. When asking for assistance, developers knowingly admit that their codebase violates specific design principles such as the existence of large modules (i.e., low cohesion) and duplicate or near-duplicate code. For instance, in Quote~\ref{Quote:Architecture_}, the developer realizes that a method in their class performs more actions than it should and seeks advice on how best to refactor the method to adhere to the single responsibility principle. Such code negatively impacts both program comprehension and unit testing. Also, similar to the other topics, there exist situations where structurally based refactoring leads to issues to which developers turn to the community for assistance. 

By following well-established architecture/design principles and patterns, developers construct their system using concrete and well-tested solutions and at the same time promote code reuse, performance, reliability, and extensibility. Additionally, the vocabulary of the pattern conveys the purpose of the pattern and thereby aids in communication between developers. %Furthermore, reviewing the list of Self-Affirmed Refactoring patterns (refer to Table \ref{Table:SAR_Patterns}), we encounter frequent occurrence of terms that are indicative of the reason developers turn to well-established patterns, such as complexity, coupling, readability, performance, and configurability. 
Finally, Table \ref{Table:LADTopicsWords} shows some of the common patterns developers mention in their post (e.g., `singleton', `factory', and `view controller').

\begin{center}
\fbox{\parbox{\dimexpr\linewidth-2\fboxsep-2\fboxrule\relax}{\centering
\textbf{Can this MVC code be refactored using a design pattern?}

\begin{flushleft}
\textit{``I've got controller code like this all over my ASP.NET MVC 3 site... we have caching, user reputation handling, auditing, all in one. Doesn't really belong in one spot does it. Hence the problem with the current code, and the problem with trying to figure out how to move it away.''}
\end{flushleft}
}}
\captionof{Quote}{Sample question for the architecture/design patterns topic where a developer requires assistance with making their code more robust by adhering to the single responsibility principle
\cite{lda_arch10}.\label{Quote:Architecture_}}
\end{center}

In summary, this topic shows that as a system evolves, the accumulation of updates made to the source code often results in the structure of the codebase violating established design principles and patterns. Developers face challenges refactoring their code to revert these violations and seek assistance from the community around applying  SOLID, DRY, SRP, and KISS principles to their codebase. For the research community, this presents an opportunity to investigate how to provide stronger advice to developers attempting to stick to best practices. One potential avenue is to look at heuristics or AI model-based approaches to pointing out violations of, for example, DRY.

% \textcolor{red}{\subsubsubsection*{\textit{\textbf{Primary Challenge}}}
% As a system evolves, the accumulation of updates made to the source code often results in the structure of the codebase violating established design principles and patterns. Developers face challenges with refactoring their code to revert these violations and seek assistance from the community around applying  SOLID, DRY, SRP, and KISS principles to their codebase.} 

% \textcolor{red}{\subsubsubsection*{\textit{\textbf{Key Takeaways}}}
% \underline{\textit{For Practitioners:}} Developers should ensure that their software process facilitates the updating and reviewing of technical design artifacts for the required updates to the source code. This approach will ensure that developers are aware of the architecture of the system and the underlying design principle they need to adhere to when updating their code.}

% \noindent\textcolor{red}{\underline{\textit{For Researchers:}} While machine learning has been utilized to determine the readability of blocks of source code, the research community should investigate the feasibility of using AI models to determine program comprehension from a higher/abstract level by analyzing the architecture/design of a system.}

%%%%%%%%%%%%%%%%%%%%%%%%%%%%%%%%%%%%%%%%%%%%%%%
\subsubsubsection{Unit Testing} \label{section:lda_testing}
% Our observation shows that the majority of these questions are related to developers having problems with writing unit tests. In some instances, we observe that developers struggle to write unit tests that accommodate the refactored production code (e.g., Quote~\ref{Quote:Test_01}). Additionally, our reviews also show developers seeking advice around creating and using mocks for the code under test (e.g., Quote~\ref{Quote:Test_02}).

% Our analysis also shows developers asking for general advice around the refactoring of test suites. This includes best practices around test-driven development involving refactored code (e.g., Quote~\ref{Quote:Test_03}), configurations to the test suite (e.g., Quote~\ref{Quote:Test_04}), and refactoring questions specific to the developer's existing test suite (e.g., Quote~\ref{Quote:Test_05}).

Unit testing is an essential part of ensuring the quality of a system. Similar to production code, developers also refactor their test cases to meet internal quality requirements and/or to support the refactored production code. In most instances, developers face challenges with writing test cases, primarily to accommodate refactored production code. Since there can be more than one test method to evaluate a single production method, an update to the structure of production code usually involves extensive updates to the test suite. For example, in Quote \ref{Quote:Test_}, the developer runs into issues with the test suite after refactoring the system under test (i.e., production code) and seeks advice from the community on the appropriate approach that needs to be followed to update test cases.

Furthermore, it should be noted that while we do encounter questions around test configuration and API usage (e.g., mocking), the majority of the questions revolve around the developer's specific project code. Finally, the presence of the terms complexity, maintainability, and testability in the Self-Affirmed Refactoring patterns list (refer to Table \ref{Table:SAR_Patterns}) shows that developers recognize the importance of writing code that is test friendly. By reducing the cohesiveness and cyclomatic complexity of their production code, developers will find it much easier to construct test cases that achieve a high degree of modularization (i.e., responsibility) and code coverage.

\begin{center}
\fbox{\parbox{\dimexpr\linewidth-2\fboxsep-2\fboxrule\relax}{\centering
\textbf{TDD and Refactoring the “system under test”}

\begin{flushleft}
\textit{``There were requirements changes and I had to change some classes behavior and API Changing one class behavior eventually led to changing a few others. I didn't know how to start this process from the test side, so I started changing the code. I ended up with lots of compilation errors in the test code and after I fixed them some did not pass. But the thing is, I don't even know if the tests cover what they used to cover before... TDD is supposed to give me a safety net while refactoring. Isn't it? As it currently appears in my case it doesn't give me that.''}
\end{flushleft}
}}
\captionof{Quote}{Sample question for the unit testing topic where a developer requires assistance updating the test suite due to production code refactoring
\cite{lda_test03}.\label{Quote:Test_}}
\end{center}

In summary, test cases, like production code, are subject to evolution during the system's lifetime. However, evolving a test suite alongside production code can be challenging, and developers seek assistance incorporating these changes into the test suite. For instance, there can be multiple test methods (i.e., test cases) associated with a single production method. Hence, the refactoring of a single production method will require multiple updates to the test suite. The research and vendor communities can support developers with tools that either give recommendations or automatically refactor test suites when developers refactor the system under test.

% \textcolor{red}{\subsubsubsection*{\textit{\textbf{Primary Challenge}}}
% Test cases, like production code, are subject to evolution during the system's lifetime. However, evolving a test suite alongside production code can be challenging, and developers seek assistance incorporating these changes into the test suite. For instance, there can be multiple test methods (i.e., test cases) are associated with a single production method. Hence, the refactoring of a single production method will require multiple updates to the test suite.}

% \textcolor{red}{\subsubsubsection*{\textit{\textbf{Key Takeaways}}}
% \underline{\textit{For Practitioners:}} As part of the system design stage in a project, the technical design artifacts should also document and take into account the design of the test suite.}

% \noindent\textcolor{red}{\underline{\textit{For Educators:}} Emphasize the importance of code quality in test suites with respect to the maintenance and evolution of the system as a whole.}

% \noindent\textcolor{red}{\underline{\textit{For Researchers/Vendors:}} Provide developers with tools that either give recommendations or automatically refactor test suites when developers refactor the system under test.}

%%%%%%%%%%%%%%%%%%%%%%%%%%%%%%%%%%%%%%%%%%%%%%%
\subsubsubsection{Database} \label{section:lda_database}
% Questions under this topic look at advice for refactoring code that deals with accessing data stores. The majority of the questions in this topic are related to the refactoring of SQL queries. From a non-functional perspective, developers ask more questions around maintainability and performance. In terms of performance, developers look for help with reducing query execution time (e.g., Quote~\ref{Quote:Database_01}) and reducing memory footprint by optimizing queries to return only the required set of records (e.g., Quote~\ref{Quote:Database_02}). From a maintainability perspective, analyzability is a key concern among developers. These concerns range from renaming database elements to the simplification of queries. When it comes to renames, developers solicit advice on performing batch/bulk rename operations (e.g., Quote~\ref{Quote:Database_03}). Under query simplification, developers require help with reducing the length of a query (e.g., Quote~\ref{Quote:Database_04}), optimization of queries containing subqueries to remove duplicate statements (e.g., Quote~\ref{Quote:Database_05}), and replacement of subqueries with joins (e.g., Quote~\ref{Quote:Database_06}).

% Finally, similar to source code, developers also aim to modularize their queries as a means to improve maintainability as in the case of this question-- Quote~\ref{Quote:Database_07}, where the developer seeks assistance from the community to combine two queries into one.

Traditionally refactoring has been primarily focused on improving the quality of program source code. However, these are not the only software engineering artifacts that developers refactor in real-world systems \cite{MensTSE2004}. One such artifact is a database. A database is a crucial part of most software systems, and a highly optimized database ensures better performance in terms of speed and resource utilization, among other attributes \cite{Ambler2006Refactoring}.To this extent, most questions in this topic revolve around the refactoring of SQL queries to resolve challenges around performance and maintainability. In terms of performance, developers look for help reducing memory and query execution time by optimizing queries to return only the required set of records. From a maintainability perspective, improving analyzability or program comprehension is a crucial concern among developers. Like optimizing source code, developers seek assistance with reducing the complexity and length of queries. As an example, in Quote~\ref{Quote:Database_}, the developer seeks assistance to improve the performance of a query and at the same time indirectly states that the complexity and readability of the query needs improving. Further, we also observe developers seeking assistance with performing batch/bulk rename operations to database elements. Finally, developers also require help improving the reusability or modularization of queries by removing duplicate code and adherence to the single responsibility principle. 

\begin{center}
\fbox{\parbox{\dimexpr\linewidth-2\fboxsep-2\fboxrule\relax}{\centering
\textbf{Performance issue - refactor select subqueries which are doing the same joins}

\begin{flushleft}
\textit{``I am getting performance issues because in the subqueries, I need to join on the same tables for each subqueries which is an heavy operation. Consider the below (ugly) example... so to me it sounds like the joins in subqueries is overkill even if it is working fine (very slow...)''}
\end{flushleft}
}}
\captionof{Quote}{Sample question for the database topic where a developer requires assistance improving the performance and maintainability of a SQL query
\cite{lda_db08}.\label{Quote:Database_}}
\end{center}

In summary, from this topic, we observe that developers prefer to implement business logic within SQL scripts and these queries tend to grow in length and complexity. As such, this negatively impacts code readability, design principles, and system performance. The research/vendor community should provide developers with tools that automatically refactor or suggest changes to database elements based on source code refactoring and vice versa. Additionally, research into the readability of SQL scripts will lead to metrics and tools that developers can utilize in their implementation workflow.

\begin{tcolorbox}
\textbf{Summary for RQ$_2$.}
Our analysis of refactoring discussions shows questions revolving around five topics-- \textit{Code Optimization}, \textit{Tools and IDEs}, \textit{Architecture and Design Patterns}, \textit{Unit Testing}, and \textit{Database}. The primary driver behind these questions is the need to improve non-functional quality attributes in the code, of which improving maintainability is a key concern. Improving readability (such as reducing lengthy conditional statements) and reusability is of utmost concern for developers and is not only related to source code. Furthermore, synchronizing refactoring changes across software engineering artifacts (such as unit tests and databases) is also challenging for developers. Additionally, we highlight takeaways under each topic for the key stakeholders.
\end{tcolorbox}

\subsection{\RQC}
\subsubsection*{\textit{Motivation \& Approach:}}
The purpose of this RQ is to understand the types of refactoring questions asked by developers that are challenging to answer. Additionally, we also look at the type of questions that the Stack Overflow community considers popular (or attractive). To this extent, this RQ builds on the results of the prior RQ to understand the specific refactoring topics that are considered popular or challenging by the community. 

Our study of popularity and difficulty of topics is similar to prior research \cite{Yang2016JCST,Bagherzadeh2019FSE,Abdellatif2020MSR}. We measure the popularity of a topic by looking at the average view count, favorite count, and score of questions associated with each topic. The higher the average value for each metric, the more popular the topic. When it comes to topic difficulty, we look at the questions that do not have any answers, do not have an accepted answer, and the median time the community takes to provide an acceptable answer to a question. Since an answer can only be set as an accepted answer by the developer who asks the question, there can be situations where the person asking the question forgets to mark a provided answer as an acceptable answer. Hence, we look at the percentage of questions with no answers and those that have accepted answers. 

\begin{table*}
\centering
\caption{Popularity and difficulty of refactoring topics.}
\label{Table:PopularityDifficulty}
\begin{adjustbox}{width=1.0\textwidth,center}
\begin{tabular}{@{}lrrr|rrrrr@{}}
\toprule
\multicolumn{1}{c}{\multirow{3}{*}{\textbf{Topic}}} & \multicolumn{3}{c|}{\textit{\textbf{Popularity Metrics}}} & \multicolumn{5}{c}{\textit{\textbf{Difficulty Metrics}}} \\ \cmidrule(l){2-9} 
\multicolumn{1}{c}{} & \multicolumn{3}{c|}{\textbf{Average Counts}} & \multicolumn{2}{c}{\textbf{\begin{tabular}[c]{@{}c@{}}Questions w/o \\ an ans.\end{tabular}}} & \multicolumn{2}{c}{\textbf{\begin{tabular}[c]{@{}c@{}}Questions w/o \\ an accepted ans.\end{tabular}}} & \multicolumn{1}{c}{\textbf{\begin{tabular}[c]{@{}c@{}}Median hrs.\\  to an \\ accepted ans.\end{tabular}}} \\
\multicolumn{1}{c}{} & \multicolumn{1}{c}{\textbf{Views}} & \multicolumn{1}{c}{\textbf{Favorites}} & \multicolumn{1}{c|}{\textbf{Score}} & \multicolumn{1}{c}{\textbf{Count}} & \multicolumn{1}{c}{\textbf{Pct.}} & \multicolumn{1}{c}{\textbf{Count}} & \multicolumn{1}{c}{\textbf{Pct.}} & \multicolumn{1}{l}{} \\ \midrule
Code Optimization & 168.2 & 1.09 & 0.81 & 277 & 6.69\% & 1,289 & 31.12\% & 0.23 \\
Tools and IDEs & 450.9 & 1.61 & 1.81 & 316 & 13.19\% & 1,014 & 42.34\% & 0.40 \\
Architecture and Design   Patterns & 247 & 1.14 & 1.29 & 122 & 7.35\% & 594 & 35.78\% & 0.32 \\
Unit Testing & 289.23 & 1.67 & 1.33 & 96 & 8.84\% & 406 & 37.38\% & 0.34 \\
Database & 218.3 & 1.26 & 1.00 & 17 & 8.95\% & 69 & 36.32\% & 0.44 \\
\textit{Refactoring Average} & \textit{274.73} & \textit{1.35} & \textit{1.25} & \textit{165.60} & \textit{9.00\%} & \textit{674.4} & \textit{36.59\%} & \textit{0.34} \\ \bottomrule
\end{tabular}
\end{adjustbox}
\end{table*}

\subsubsection*{\textit{Findings:}}
\subsubsection{Popularity}
From Table \ref{Table:PopularityDifficulty}, we see that the topic \textit{Tools and IDEs} is the most popular among the five refactoring topics while \textit{Database} is the least popular topic. Even though the total number of \textit{Tools and IDEs} questions are less than \textit{Code Optimization} questions, the average view count metrics of the former are higher than the latter (by a percentage difference of approximately 91.33\%). This indicates that developers are more frequently searching for, and thereby viewing, questions around refactoring tools and IDEs, perhaps, looking for help with a similar problem they are experiencing.  

Additionally, we also look at the same metrics for all non-refactoring Stack Overflow questions (i.e., questions that are not part of our refactoring dataset) with the goal of comparing the popularity of refactoring questions against all other types of questions. In general, non-refactoring Stack Overflow questions have an average view count of  2361.42, an average favorite count of 0.62, and an average score of 2.08. Looking at the two sets of values, we see that refactoring questions have more favorites than general questions. At the same time, the view count is much higher for general Stack Overflow questions than refactoring questions (a percentage difference of approximately 158.313\%). The score metric for both these two types of questions is similar. 

\subsubsection{Difficulty}
Once more, looking at Table \ref{Table:PopularityDifficulty}, we observe that the topic \textit{Tools and IDEs} has the most number of questions without both an accepted answer or any answer for that matter, and takes around 0.40 hours to receive an accepted answer. Hence, this seems to be the most challenging type of question for developers to answer. Topics falling under \textit{Code Optimization} are less challenging to answer, as only 6.69\% of these questions do not have an answer, while 31.12\% of the questions do not have an accepted answer. Furthermore, it takes around 0.23 hours for such questions to receive an accepted answer.

\subsubsection{Popularity \& Difficulty Correlation}
Similar to the prior work mentioned above, we perform a correlation analysis of the three popularity metrics (i.e., average view, favorite, and score)  against the three difficulty metrics (median time to obtain an accepted answer, percentage of questions without any answers, and without an accepted answer). A Shapiro-Wilk normality test \cite{Taeger2014Statistical} on these variables shows that the data follows a normal distribution; therefore, we Pearson correlation test \cite{Taeger2014Statistical}. Table \ref{Table:PopularityDifficultyCorrelation} shows the results of our correlation analysis. This table shows a strong positive statistically significant correlation (i.e., $p$-value $<$ 0.05) for the difficulty metric percentage of questions without an accepted answer and the popularity metrics average views and score. Additionally, there is a strong positive relationship between views and posts without any answers. The remaining popularity and difficulty metrics do not show any statistically significant correlations. From this, we see that questions without an accepted answer are considered to be difficult, yet interesting enough that they garner views and scores from the developer community. This phenomenon is observable with the topic metrics in Table \ref{Table:PopularityDifficulty}.

\begin{table}
\centering
\caption{Pearson correlation analysis between the popularity and difficulty of refactoring topics.Bold values indicate a statistically significant correlation (i.e., $p$-value $<$ 0.05).}
\label{Table:PopularityDifficultyCorrelation}
\begin{tabular}{@{}lrrr@{}}
\toprule
\multicolumn{1}{c}{\multirow{2}{*}{\textit{coefficient / $p$-value}}} & \multicolumn{3}{c}{\textbf{Average}} \\
\multicolumn{1}{c}{} & \multicolumn{1}{c}{\textbf{Views}} & \multicolumn{1}{c}{\textbf{Favorites}} & \multicolumn{1}{c}{\textbf{Score}} \\ \midrule

\textbf{\% w/o any answer} & {\textbf{0.921/0.026}} & 0.753/0.142 & 0.816/0.092 \\
\textbf{\% w/o accepted answer} & {\textbf{0.950/0.013}} & 0.750/0.144 & {\textbf{0.949/0.014}} \\
\textbf{Hrs. to accepted answer} & 0.453/0.443 & 0.424/0.477 & 0.423/0.478 \\ \bottomrule

% \textbf{\% w/o any answer} & 0.738/0.155 & 0.791/0.111 & 0.738/0.155 \\
% \textbf{\% w/o accepted answer} & {\textbf{0.975/0.005}} & 0.872/0.054 & {\textbf{0.975/0.005}} \\
% \textbf{Hrs. to accepted answer} & 0.400/0.517 & 0.600/0.350 & 0.400/0.517 \\ \bottomrule
\end{tabular}
\end{table}

\subsubsection{Unanswered Questions}
In our final analysis, we examine unanswered questions (i.e., questions without an accepted and non-accepted answer post). To this extent, two of the authors manually reviewed a stratified statistically significant sample of 259 unanswered questions. This sample represents a confidence level of 95\% and an interval of 10\% for each topic, from a total of 784 unanswered questions. As part of the review, the authors examined the unanswered questions for ambiguity, incompleteness, or lack of concrete examples (source code, diagrams, etc.) to determine the lack of developer interaction with these questions. %\hl{Based on the manual analysis, only 20\% did not contain code snippets, and about 15\% were incomplete or/and confusing to the authors.}\anthony{need to double check these numbers}

The examination of these questions reveals that a majority were not ignored per se but received comments instead of answer posts. The content of these comments, in most situations, provides the author of the question with high-level suggestions on addressing the question or requests more clarification about the question. Since the volume of content permitted in a comment is restricted when compared to an answer post\footnote{\url{https://meta.stackexchange.com/questions/19756}}, most suggestions in comments were brief and tend to provide hyperlinks to other resources such as other Stack Overflow questions/answers, API documentation, and blog posts (e.g., \cite{unanswered04}). With regards to clarifications, we observe developers utilizing comments as a means to have a back-and-forth discussion for clarity on the problem faced by the developer (e.g., \cite{unanswered01}). Additionally, in some comments, the respondents simply state that there is no solution for the developer's issue (e.g., \cite{unanswered03}) or is a known or reported bug (e.g., \cite{unanswered08}). We also observe that a minority of unanswered questions are answered by the developer asking the question; the answer either appears as a comment or as an edit to the question (e.g., \cite{unanswered02}). Finally, it was interesting to note that there exist some unanswered questions that were not refactoring related (i.e., the developer misuses the refactoring term or tag in the question), showing that most of the developers understand the purpose of refactoring and its application (e.g., \cite{unanswered05}). Hence, the majority of unanswered questions were not actually ignored by the community but responded to through comments. %, as the responders choose to comment when they are not giving a definitive answer, or requiring the enquirer to check other resources.

Examining the questions' body, we observe that most of the questions around tools were asking for help in solving issues specific to the developer's project, such as altering the standard find-and-replace or refactoring operation for a particular purpose (e.g., \cite{unanswered05}). We also observe that most questions around improving or implementing code reusability are often focused on identifying best practices \cite{unanswered07}, which is usually considered an opinion-based question by the Stack Overflow community. Such questions are often discouraged by the community as there is no commonly accepted answer and thus end up going unanswered even if they are comparatively coherent (i.e., it is clear what the author's problem and intent were). At times, such questions are suggested to be migrated to another site, most commonly Code Review Stack Exchange\footnote{\url{https://codereview.stackexchange.com}}, an affiliated site, to be better addressed. %Improvements also focus on structural optimization.

\begin{tcolorbox}
\textbf{Summary for RQ$_3$.}
Questions around refactoring tools/IDEs are popular and challenging to answer, while questions around optimization of code snippets are the least difficult to answer. Questions that do not have an (accepted) answer usually have responses from the community as comments, which are usually high-level suggestions to address the issue or a request for clarification.
% Questions around refactoring based tools and IDEs gain the most interest among the developer community. These questions are not only popular but are also challenging to answer. On the other hand, questions soliciting advice around the optimization of code snippets are the least difficult to answer, but their popularity is far less than tool based questions. We also observe a statistically significant correlation between the percentage of questions without an accepted answer and the average views and score of the questions. Finally, we observe that most unanswered questions contain responses from the community in the form of comments. These responses are mostly high-level suggestions to address the issue or request for further clarification.
\end{tcolorbox}

\section{Discussion and Takeaways} 
\label{Section:Discussion}
Our findings show that Stack Overflow is a popular venue for developers to seek assistance with refactoring challenges for various technologies. Developers can post refactoring questions related to a range of technologies and artifacts, and usually receive a response in a short period of time. By performing a manual analysis of a statistically significant set of question posts, in our RQs, we supplement our quantitative findings and obtain an accurate understanding of the challenges developers face when refactoring their systems. Furthermore, our findings empower educators to update their course curriculum to reflect real-world settings better. For instance, 1) instilling the need for students to practice test-driven development in projects, 2) the importance of conducting early and frequent reviews of all types of software engineering artifacts, and 3) ensuring that these artifacts capture the non-functional goals of the system (especially around readability and reusability). In this section, through a series of takeaways, we discuss how our findings support the community.

\subsubsection*{\textbf{Research Community Takeaways}}
While the research community has made considerable strides in refactoring related research, our findings demonstrate the challenges developers face in real-world projects. These findings highlight the gaps between the academic definition of refactoring and its actual usage in real-world settings. Furthermore, they provide the research community with opportunities to further evolve the field.

\vspace{3.0mm}\noindent\textit{\underline{Adaptation of refactoring operations for multiple programming language and artifact types}}
\vspace{1.5mm} 

\noindent Researchers have traditionally based their refactoring studies on statically typed programming languages (especially Java). However, our findings show that while this does benefit developers in real-world scenarios, there are opportunities for researchers to evolve the field further and increase the diversity of their research. To this extent, there is a need to adapt traditional refactoring operations to support dynamic language types (e.g., Python and JavaScript), which are rising in popularity. Further, the research community should also examine the possibility of deriving refactoring operations specific to dynamic languages. Additionally, programming source code files are not the only artifacts that developers associate with quality. Our findings show challenges with improving the quality around other related artifacts such as database elements (tables, columns, queries, etc.) and test suites, which are not frequently studied in research—further highlighting opportunities to evolve the refactoring field.

\vspace{3.0mm}\noindent\textit{\underline{Improve and extend the applicability of readability quality metrics}}
\vspace{1.5mm} 

\noindent Our findings show that improvements to readability are a critical concern for developers. While the research community has made considerable strides in producing readability metrics and models \cite{buse2009learning,scalabrino2018comprehensive}, the community needs to better collaborate with established vendors in integrating their contributions with popular tools and IDEs to promote the usage of their artifacts. Additionally, our findings highlight specific avenues for readability research, such as optimizing/eliminating lengthy switch-case statements and conditional loops and understanding their influence on comprehension. Our findings also show that developers seek assistance with improving the readability of database artifacts, such as SQL queries and table/column names. While database vendors and the research community have provided developers with material to optimize the setup and performance of database systems, there is not much support around readability improvements to database elements. Developers specifically struggle with performing bulk renaming of database elements and improving the readability of long and complex queries.

\vspace{3.0mm}\noindent\textit{Expand the study and applicability of reusability beyond source code}
\vspace{1.5mm} 

\noindent Along the lines of readability, developers also seek assistance with improving reusability, and like readability, the reusability assistance is not limited to source code. In addition to removing duplicate code from source code, developers also seek assistance with improving the reusability of database queries. To this extent, the research community should investigate refactoring at an architectural level to provide developers with information and recommendations around the structure of their codebase to improve reusability. Furthermore, there is also an opportunity to conduct research around reusability metrics and models for database artifacts.

\subsubsection*{\textbf{Tool/IDE Vendor Community Takeaways}}
Like the research community, tool and IDE vendors play a vital part in ensuring developers write and maintain quality code. Hence, there are specific findings from our study which vendors can utilize to enhance their tools/IDEs. Our study shows that developers either mention the IDE/tool when providing context to their problem or ask questions specifically around the IDE/tool. 

\vspace{3.0mm}\noindent\textit{\underline{Automatic synchronization between project artifacts.}}
\vspace{1.5mm} 

\noindent We observe a trend that while source code is central to refactoring, developers struggle with updating other artifacts (e.g., test cases, databases) due to refactoring of the source code. For instance, a single production method can be evaluated by more than one test method. Hence, refactoring of production code can result in multiple updates to the test suite. Likewise, renames to database tables/columns should cascade to SQL scripts and the data access layer in the source code. To this extent, IDE's should provide users with the ability to either synchronize refactorings across artifacts automatically or at the very least recommend refactoring opportunities in the related artifacts.

\vspace{3.0mm}\noindent\textit{\underline{Enhanced rename refactoring functionality.}}
\vspace{1.5mm} 

\noindent Our findings show that as most questions revolve around renames, IDE/tool vendors should consider incorporating the rename refactoring work by Peruma et al. \cite{Peruma2020JSS,Peruma2021ICPC}, Liu et al. \cite{Liu2015TSE}, Arnaoudova et al. \cite{Arnaoudova2014TSE}, and Allamanis et al. \cite{Allamanis2014FSE} into their products to better provide developers with an automated approach to identifying, appraising and suggesting high-quality identifier names.

\vspace{3.0mm}\noindent\textit{\underline{Enhance the user experience.}}
\vspace{1.5mm} 

\noindent In addition to providing extensive and innovative refactoring functionality in their tools/IDEs, vendors must ensure that their products also exhibit an optimal user experience. Usability and trustworthiness are an essential part of refactoring tool adoption and are among the reasons for the lack of usage \cite{Murphy-Hill2012TSE,Eilertsen2021SANER}. Our study corroborates these findings where we observe developers requiring help to configure tools or finding a tool for a specific purpose.

\subsubsection*{\textbf{Developer Community Takeaways}}
Developers can also utilize our findings to ensure they follow a disciplined approach to software implementation. Our findings can be incorporated into the project's software development process as checklists/guidance that developers need to adhere to before certification of a release/deployment. Organizational managers can also utilize our findings to ensure that their development team is trained in the necessary skills required for refactoring and has access to the appropriate tools.

\vspace{3.0mm}\noindent\textit{\underline{Extend coding standards utilized in projects to support naming standards for all project artifacts.}}
\vspace{1.5mm} 

\noindent While it is common for project teams to utilize organizational or technology-specific coding standards, teams should ensure that the standards in use apply to all types of project artifacts utilized in the project. For instance, while general standards define the naming standards for source code identifier names (e.g., method names should begin with a verb), they do not define the naming standards for database table/column names. Furthermore, teams should also be made aware of the concept of linguistic anti-patterns \cite{Arnaoudova2016EMSE} and how to detect and correct such occurrences in the code \cite{Peruma2021ICMSE}.

\vspace{3.0mm}\noindent\textit{\underline{Integrating code quality tools into the build process for the early detection of poor coding practices.}}
\vspace{1.5mm} 

\noindent Using code quality tools (e.g., code/design smell detectors \cite{Aljedaani2021EASE}) during implementation will supplement the review tasks by automating the time-consuming task of detecting poor-programming practices (e.g., duplicate code, poor identifier naming, high cyclomatic complexity). Furthermore, there should be necessary checks in the project's process to ensure that developers can only ignore specific quality rules after providing valid justification.

\vspace{3.0mm}\noindent\textit{\underline{Perform frequent and early peer-reviews on all project artifacts.}}
\vspace{1.5mm} 

\noindent Peer reviews should not be limited to source code. Other artifacts such as architecture/design documents, test suites, and database artifacts should also undergo peer-reviews. Early reviews of such artifacts will ensure that the implemented system meets its non-functional goals. For instance, code reviews will help address readability issues before it accumulates to be a serious concern. Architectural and design artifacts will ensure that the system follows appropriate design principles, such as addressing reusability and modularization. Likewise, reviews of database queries will help in addressing readability and performance issues. Furthermore, these database-related reviews will also help to ensure the extent to which business logic is contained within SQL scripts (i.e., stored procedures) versus in the source code.

\section{Threats To Validity}
\label{Section:Threats}
This section discusses the threats that may potentially impact the validity of our study. We group the treats into three categories-- Internal, External, and Construct \cite{Wohlin2012Experimentation}. 

\smallskip
\noindent\textbf{Internal Validity:} \textit{These are factors that influence our results}. We constructed our dataset by extracting and analyzing questions with the `refactor' tag or contain the term `refactor' in the title. There is the possibility that we may have excluded synonymous terms/phrases. However, even though this approach reduces the number of posts in our dataset, it also decreases false positives. Our approach ensures that we analyze posts that are explicitly geared towards refactoring challenges faced by developers. In other words, these are posts where developers were explicitly considering a refactoring action and were aware that they were attempting refactoring. Additionally, as the goal of this study is to understand the refactoring challenges faced by developers, our analysis is focused on the questions posted by developers. As with similar studies, our study is limited to analyzing only the most recent version of a post. Our analysis also does not take into account comments associated with a post, as comments are not considered as answers; comments in Stack Overflow are considered as temporary ``Post-It'' notes, and not every user has the privilege of creating a comment\footnote{\url{https://stackoverflow.com/help/privileges/comment}}.

\smallskip
\noindent\textbf{External Validity:} \textit{These are factors that impact the generalizability of our findings}. Even though there are several technology-based question and answer websites, our scope (and analysis) focuses exclusively on Stack Overflow-- the largest such site on the Stack Exchange network\footnote{\url{https://data.stackexchange.com/}} that caters to a wide range of computer programming topics. Furthermore, from RQ$_1$ (Section \ref{SubSection:RQA}), we observe a large quantity of refactoring question and answer posts asked and answered by a diverse set of developers relating to various technologies. Additionally, the \textit{SOTorrent} dataset, containing the Stack Overflow dump, has been widely used in similar knowledge sharing based studies. While we recognize that developer surveys/interviews are also viable mechanisms to study refactoring challenges, our study captures questions posted by 7,795 distinct Stack Overflow users. Furthermore, our findings provide either a starting or comparison point for future participatory-based research. Finally, while it is true that our findings capture the state of refactoring at the time we conducted our study, our results present snapshots of the data, which future studies can leverage to examine (e.g., via replication-based studies) the evolution of the field and also the state of Stack Overflow.

\smallskip
\noindent\textbf{Construct Validity:} \textit{Here we identify the extent to which our experiments are designed to measure what they are supposed to measure}. For RQ's that involves a qualitative analysis (such as annotations), we manually analyze a sample set of posts. However, these samples are statistically significant and followed a peer-review process to counter any bias in annotating. Further, since the review process involved a discussion between the annotators for each conflicting annotation, there was no need for calculating the inter-annotator agreement as the finalized dataset was conflict-free. The use of LDA in our topic modeling algorithm can be considered as a threat. However, as mentioned in the RQ (Section \ref{SubSubSection:RQBC}), this algorithm has been heavily utilized in similar studies. Furthermore, our selection of five topics is based on our analysis of fifty topics (in increments of one); our analysis included evaluating topic coherence, perplexity score, and visualization. Additionally, we also manually review a statistically significant sample of questions associated with each of the five topics. Finally, even though we utilize the Pearson correlation coefficient to measure the relationship between variables, there are other statistical measures, such as Cohen's $d$ and ANOVA, which can also be applied to the data.

\section{Conclusion and Future Work}
\label{Section:Conclusion}
Software refactoring is an essential activity in the maintenance and evolution of software. However, given the complexity of a system and the experience of the developer maintaining the system, performing refactoring operations can prove to be challenging. Hence, developers usually seek assistance from the community through question-and-answer websites such as Stack Overflow.

In this empirical study, we perform a quantitative and qualitative analysis of refactoring questions asked by developers on Stack Overflow. Our quantitative approach involved applying statistical measures on the mined data, while the qualitative analysis involved a manual review of a statistically significant sample of questions. Our results show that Stack Overflow is a popular online resource for developers to seek assistance with refactoring challenges. Our findings show that while most developers seek assistance with traditional statically typed languages (specifically Java and C\#), there is a growing increase in refactoring dynamically typed code such as Python and JavaScript. Looking at the topics developers need assistance with, we observe that most questions are around optimizing source code to improve readability and reusability. However, source code is not the only artifact that developers refactor. Other artifacts, specifically database-related elements, are also subject to refactoring. Furthermore, developers find it challenging to propagate refactoring changes between related project artifacts. Tools are also a popular discussion topic among developers, specifically around the renaming of content within non-source code files and advanced refactoring automation. From our findings, we highlight a series of actionable takeaways for relevant stakeholders that will evolve the field of refactoring and improve developer productivity.

For our future work, we plan on conducting a structured survey with both junior and senior software developers from both open-source and industry. The survey will explore their general and specific challenges when performing refactoring activities; this includes (but is not limited to) software engineering artifacts, tools, and technologies associated with refactoring activities. This survey will complement and validate our current Stack Overflow study to provide the software engineering community with a more comprehensive view of refactoring practices.
% For our future work, we plan on studying the answers associated with refactoring questions to understand how developers respond to refactoring questions, including the refactoring terminology in use and the common suggestions to refactoring challenges. Additionally, we plan on investigating the differences between an accepted answer and a non-accepted answer. We also plan to conduct a structured survey with both junior and senior software developers from both open-source and industry to learn about their general and specific challenges when performing refactoring activities. This survey will complement our current Stack Overflow study to provide the software engineering community with a more comprehensive view of refactoring practices.

\section{Acknowledgments}
\label{Section:Acknowledgment}
We would like to thank the reviewers at ESE for their detailed and invaluable feedback.
\bibliographystyle{spmpsci}      % mathematics and physical sciences

\bibliography{references.bib}   % name your BibTeX data base

\end{document}